\documentclass[12pt,preprint]{aastex}

\slugcomment{Revised Nov 1, 2004}

\newcommand\lam{$\lambda$}
\newcommand{\kms}{km~s$^{-1}$}
\newcommand\etal{et~al.}

\newcommand\CaII{Ca\,{\sc ii}}
\newcommand\CII{C\,{\sc ii}}
\newcommand\CIIstar{C\,{\sc ii}*}
\newcommand\CIV{C\,{\sc iv}}
\newcommand\FeIII{Fe\,{\sc iii}}
\newcommand\FeIX{Fe\,{\sc ix}}
\newcommand\FeX{Fe\,{\sc x}}
\newcommand\FeXII{Fe\,{\sc xii}}
\newcommand\HI{H\,{\sc i}}
\newcommand\MgI{Mg\,{\sc i}}
\newcommand\MgII{Mg\,{\sc ii}}

\newcommand\NaI{Na\,{\sc i}}
\newcommand\OI{O\,{\sc i}}
\newcommand\OIV{O\,{\sc iv}}
\newcommand\OV{O\,{\sc v}}
\newcommand\OVI{O\,{\sc vi}}
\newcommand\OVII{O\,{\sc vii}}
\newcommand\PIV{P\,{\sc iv}}
\newcommand\SiIV{Si\,{\sc iv}}

\begin{document}

\title{A Survey of \OVI\ Absorption in the Local Interstellar Medium}
\author{William R. Oegerle\altaffilmark{1}, 
Edward B. Jenkins\altaffilmark{2},
Robin L. Shelton\altaffilmark{3}, \\
David V. Bowen\altaffilmark{2}, 
Pierre Chayer\altaffilmark{4,5}
}

\altaffiltext{1}{Laboratory for Astronomy and Solar Physics, Code 681, 
Goddard Space Flight Center, Greenbelt, MD 20771}
\altaffiltext{2}{Department of Astrophysical Sciences, Princeton University,
Princeton, NJ 08544-1001}
\altaffiltext{2}{Department of Physics and Astronomy, University of 
Georgia, Athens, GA 30602-2451}
\altaffiltext{4}{Department of Physics and Astronomy, Johns Hopkins University,
Baltimore, MD 21218}
\altaffiltext{5}{Primary affiliation: Department of Physics and Astronomy,
University of Victoria, P.O. Box 3055, Victoria, BC, V8W 3P6, Canada}

\begin{abstract}

We report the results of a survey of \OVI\ \lam1032 absorption along the
lines of sight to 25 white dwarfs in the local interstellar medium
(LISM) obtained with the {\it Far Ultraviolet Spectroscopic Explorer (FUSE)}.  
We find that interstellar \OVI\ absorption along all
sightlines is generally weak, and in a number of cases, completely
absent.  No \OVI\ absorption was detected with significance greater than
2 $\sigma$ for 12 of the 25 stars, where the 1 $\sigma$ uncertainty is
$\sim 4$ m\AA\/, equivalent to an \OVI\ column  density of $\sim 3
\times 10^{12}$ cm$^{-2}$.  Of the remaining stars, most have column
densities N(\OVI\/)$ < 10^{13}$ cm$^{-2}$ and no column densities exceed
$1.7 \times 10^{13}$ cm$^{-2}$.  For lines of sight to hot 
(T$_{\rm{eff}}>40,000$~K)  white dwarfs, there is some evidence that the
\OVI\ absorption may be at least partially photospheric or circumstellar
in origin. We interpret the ``patchy'' distribution of \OVI\ absorption
in terms of a model where \OVI\ is formed in evaporative interfaces
between cool clouds and the hot, diffuse gas in the Local Bubble (LB).
If the clouds contain tangled or tangential magnetic fields, then
thermal conduction will be quenched over most of the cloud surface, and
\OVI\ will be formed only in local ``patches'' where  conduction is
allowed to operate. We find an  average \OVI\ space density in the LISM
of $2.4  \times 10^{-8}$ cm$^{-3}$, which is similar to, or slightly
larger than, the value in the Galactic disk over kpc scales.  This local
density implies an average \OVI\ column density of $\sim 7 \times
10^{12}$ cm$^{-2}$ over a path length of 100 pc within the LB. The \OVI\
data presented here appears to be inconsistent with the model proposed
by \citet{breit94}, in which highly ionized gas at low kinetic
temperature ($\sim 50,000$~K) permeates the LB.  Our survey results are
consistent with the supernova-driven cavity picture of
\citet{coxsmith74} for the LB, and in particular, the recent model by
\citet{ch03} for the creation of cool clouds in the LB by  magnetic flux
tubes, and their subsequent magnetic shielding from conduction. 

\end{abstract}


\section{Introduction}

It has now been several decades since two important observations
revolutionized our understanding of the structure and ionization of the
interstellar medium (ISM). The first was the discovery \citep{bowyer68}
and  subsequent  investigation \citep{mccammon83} of the 0.25 keV soft
X-ray diffuse background,  which implied that a substantial fraction of
the Galactic disk in the solar neighborhood was occupied by low density,
hot ($T \ga 10^6$~K) gas.  The second was the discovery of \OVI\
absorption lines arising in the ISM \citep{rogerson73, york74, jm74,
jenkins78a}.  With an ionization potential of 114 eV,  \OVI\ is very
difficult to produce by photoionization.  In addition, the observed 
\OVI\ line profiles are broader than interstellar lines of lower
ionization potential.   Consequently, the \OVI\ ionization was most
naturally explained by collisional ionization in a hot gas  ($T \ga 2
\times 10^5$~K).  The \OVI\ line profiles have widths consistent with
thermal broadening from such a hot gas. 

On the theoretical side, \citet{coxsmith74} studied the role of
supernova explosions in creating large, low-density cavities of hot gas
in the ISM.  \citet{mo77} provided a self-consistent theory of the ISM
in which warm ($T\sim 10^4$~K) ionized clouds exist in this hot gas.  The
\citet{mo77} model attempted to explain the pressure of  interstellar
clouds, the soft X-ray background, the \OVI\ absorption, and the
ionization and velocities of the clouds.  Although numerous theoretical
models have since been proposed which emphasize the roles of magnetic
fields, or different values for the filling factor of the hot gas 
\citep{slavin92}, the general picture of a turbulent, three-phase ISM
which is churned up by supernovae has endured.  Much recent progress
has been made in developing multi-dimensional magneto-hydrodynamic simulations
of the ISM \citep{avillez00, avillez04}, which show the growth of
structures both large and small due to shocks, radiative cooling, 
and thermal instabilities.  

The origin of the \OVI\ absorption has been the subject of much debate
and modeling.  The most widely held view is that the \OVI\ is formed in
conductive interfaces between the hot, $10^6$~K gas and cool embedded
clouds \citep{mo77,cowie79}.  This explanation is consistent with a
number of characteristics of the \OVI\ absorption, such as their line
widths and kinematics. \citet{jenkins78b} has shown that the rms
dispersion in \OVI\ components is only $\sim 26$ \kms\/.  This narrow
velocity spread seems incompatible with formation in turbulent,
expanding supernova remnants.  Furthermore, \citet{cowie79} have shown
that there is a correlation between the kinematics of the \OVI\
absorption and UV lines from lower ionization lines, implying a physical
correlation between the hot and cool gas.  Finally, models of the
conductive interfaces predict column densities of \OVI\ which are in
general agreement with observations \citep{slavin89, borkowski90}. 

Despite these successes, alternative models have received a lot of
attention.  \citet{coxsmith74}, \citet{slavin92} and \citet{shelton98} 
have investigated the production of \OVI\ in the interior of bubbles
carved out in the ISM by supernova explosions.  \citet{slavin92} have
modeled the evolution of supernova remnants, including non-equilibrium
effects and magnetic fields.  They find that hot gas inside the bubble
condenses onto the interior shell wall, and will have significant column
densities of \OVI\/.  At later times in the remnant evolution ($\sim 3
\times 10^6$ yr), there is considerable cooling inside the bubble, and
large amounts of \OVI\ are produced in the middle of the bubble. In
these later phases of the bubble evolution, expansion has ceased and
consequently the \OVI\ line widths would be much narrower than in the
turbulent expanding phase \citep{shelton98}. Averaged over the bubble
lifetime, \citet{slavin92} predict the most probable value of the \OVI\
column density from one crossing of the bubble wall to be $4.5 \times
10^{13}$ cm$^{-2}$, which is higher than observed. 

Comparisons of models with observation have been hampered by the
necessarily complicated sight lines observed with {\it Copernicus}, the
only far-ultraviolet spectrograph prior to the {\it Far Ultraviolet
Spectroscopic Explorer (FUSE)} with high enough
spectral resolution to study \OVI\/ absorption.  Due to its low
effective area, however, {\it Copernicus} could only observe bright,
early-type stars that are, with a few exceptions, at distances exceeding
100 pc from the Sun. \citet{sc94} reanalyzed the {\it Copernicus} data
and showed that a significant fraction of the \OVI\ column density along
most distant lines of sight was  contributed by a region of hot gas
surrounding the Sun called the Local Bubble (LB). However, the {\it
Copernicus} data did not allow the determination of where in the LB 
the \OVI\ was formed; i.e. on conductive interfaces or in condensation
regions at the bubble wall.  The much higher sensitivity of {\it FUSE} permits
observation of nearby white dwarfs, where the problem of averaging and
blending of many different physical conditions along the line of sight
are minimized, allowing the study of the distribution of \OVI\ in the
LB.  A separate study of low-ionization species in the LISM has 
been carried out by \citet{lehner03} using many of the same lines
of sight as the present study.

\subsection{The Local Interstellar Medium}

The Sun lies inside a low density, warm cloud ($n  \sim 0.1$ cm$^{-3}$,
$T \sim 7000$~K) which extends $1-10$ pc from the Sun \citep{lb92,
lallement95}.  The Sun is located near the edge of this local
interstellar cloud (LIC) \citep{linsky00}, which is immersed inside a
hot ($T=10^6$~K), very low density ($n \sim 0.005$ cm$^{-3}$) cavity
called the Local Bubble \citep{cr87}.  Although it is unclear how
representative the LB is of other regions within our Galaxy, it provides
a unique local laboratory for studying the formation of \OVI\ in either
diffuse hot gas or in conductive interfaces to cool clouds.  Within the
LB, there are numerous clouds of low column density, although the volume
filling factor of clouds is quite small.  In addition to the LIC,
\citet{lb92} have identified the G cloud, and \citet{linsky00} have
provided evidence for two additional clouds that they dub the North and
South Galactic Pole clouds. \citet{gry95} have identified six clouds
along the line of sight to $\epsilon$ CMa, which is 132 pc distant. 
\citet{ferlet99} has estimated that the LB could contain more than 2000
diffuse clouds. Therefore, lines of sight to nearby ($50-100$ pc) white
dwarfs should pass through a minimum of one cloud interface (the LIC),
and will likely intercept several clouds.  \citet{ferlet99} notes
that most stars within only 20 pc of the Sun exhibit two or three
absorption components in \CaII\/.

It has been assumed that the LB is full of million degree gas because of
the observed emission of soft x-rays \citep{snowden98}.  The exact
origin of the LB is not known, although it is presumably due to  one or
more supernova explosions.  \citet{maiz01} and \citet{frisch98} have
presented kinematical evidence that subgroups of stars in the Sco-Cen OB
assocation passed through the current boundary of the LB about 5 million
years ago, and OB stars in these subgroups provided the supernovae that
created the LB. \citet{bb02} arrive at similar conclusions from a study
of the trajectories of moving stellar groups in the solar neighborhood,
concluding that the Local Bubble was created by up to 20 supernovae 
in the past $10-20$ Myr. \citet{sc01} also suggested that supernova explosions
in the Sco-Cen region could have resulted in  runaway OB stars that
passed close to our Sun.   A sequence of about 3 supernovae over a few
million year period from these runaway stars could provide the
energetics to power the LB.  

\citet{cox98} suggested that a substantial amount of the soft x-ray 
background might be due to charge transfer of heavy solar wind ions with
interstellar neutrals in the heliosphere (in a process analogous to the
proposed explanation of x-ray emission from comets  by
\citet{cravens00}).  This process would lead to time-dependent x-ray
emission due to fluctuations in the solar wind.  Indeed, 
\citet{snowden94, snowden98} has reported time-variable ``long-term
enhancements'' of order 30\% in the two lowest energy bins in the
ROSAT/PSPC x-ray background.  Current estimates of the contribution of
this process to the soft x-ray background range from $\sim 25$\%
\citep{snowden04} to $\sim 25-50$\% \citep{robertson01}.

\citet{vallerga96} used EUVE data to show that He is more ionized than H
in the LISM.  In subsequent work, \citet{vallerga98} argued that stellar
EUV sources are not capable of providing the observed He ionization
level.  \citet{slavin02} have modeled the soft x-ray  radiation field
within the LB near the Sun as a combination of three sources: (1)
radiation from nearby B stars and white dwarfs, (2) emission from a
$10^6$ K gas in collisional ionization equilibrium, and (3) radiation
from an evaporative interface  at the boundary of the LIC.  This model
successfully explains a number of observations, such as the ionization
of He and H, and the ionization ratios of \MgI\ to \MgII\ and
\CIIstar\/ to \CII\/ towards $\epsilon$ CMa.  Furthermore, in order to match
the observations toward $\epsilon$ CMa, an EUV radiation field larger than
that from nearby stars plus diffuse emission from the hot gas in the LB
was required, providing indirect evidence for an evaportive boundary
around the LIC.  The existence of this evaporative boundary could
be tested through observations of \OVI\/ towards nearby stars.

\citet{breit94} proposed an innovative model for the soft X-ray
background in which supernovae drive an adiabatic expansion of hot gas. 
The gas cools quickly as it expands, and highly ionized stages are
``frozen in'' to the gas.  The soft X-ray emission is due to delayed
recombination at low kinetic temperatures ($\sim 4\times 10^4$~K).  In
this model, a significant amount of \OVI\ is also produced in
wide-spread, highly excited gas within the supernova remnant.  One of
the successes of this model is its natural explanation of how cool
clouds can survive inside the LB.  If the LB contained million degree
gas, then conduction at cloud surfaces would evaporate the clouds  in a
relatively short time.  The existence of cool clouds in the LB implies 
that either the ambient medium is cool, that conduction is inoperative,
or cool clouds are being replenished.  We note that \citet{shelton03}  has
recently shown that the $\sim2\sigma$ upper limit on the intensity of
observed \OVI\ emission from a null detection along one line of sight
in the LB is about twice the intensity expected from this model.  

\citet{hurwitz04} report that a
spectrum of the LB obtained with the
Cosmic Hot Interstellar Plasma Spectrometer (CHIPS) indicates that
the diffuse EUV emission is much fainter than that predicted from
a $10^6$~K gas in collisional ionization equilibrium, as predicted
by the soft-xray observations.  The emission
measure of the \FeIX\ 171 \AA\/ line is an order of magnitude less
than predicted, and the emission lines of \FeX\ -- \FeXII\ were not
detected at all, even though they would be expected to have similar
intensities.  Depletion of iron in the LISM and/or foreground absorption could
explain the low intensity of the \FeIX\ emission, but not the
absence of the \FeX\ line.
Other possible explanations for unexpectedly low EUV emission include
non-equilibrium ionization and solar wind charge exchange effects.

\citet{ch03} have proposed a model for the origin of the
diffuse clouds in the Local Bubble, in which a magnetic flux tube
anchored at two points on opposite sides of the LB pulls free of the
boundary wall and, through its own tension, is pulled to the center of
the bubble.  The authors have investigated the dynamics of this flux
tube, and they conclude that a converging flow of matter occurs in the tube,
and leads to the formation of one or more clouds near the center of the
bubble. 

A thorough review of the LISM is not possible in this paper, and we
have just touched on subjects relevant to our survey of \OVI\ absorption.
The interested reader is referred to a recent review article on
Local Bubble research by \citet{bc04} for a more extensive discussion. 

In the next section, we describe our survey of \OVI\
absorption toward white dwarf stars within and near the LB, report the
measurements and their uncertainties.  In the following sections, we
discuss the implications of these results in terms of the models 
discussed above.  

\section{Observations and Results}

\subsection{The Survey Sample}

In this paper, we report observations of 29 nearby white dwarfs obtained
with the {\it FUSE}.  Observations used in this survey were obtained for
a variety of purposes (photometric calibration, deuterium snapshot
survey, etc), and hence the survey data are not homogeneous in terms of
signal-to-noise.  In the second year of {\it FUSE} observations, we
added white dwarfs to the Guaranteed Time Observer's target list
(program P204) in order to provide even sky coverage.   DA white dwarfs
with surface temperatures less than 40,000~K are the preferred targets,
since most of them have essentially pure hydrogen atmospheres, although
it should be noted that some percentage of DA white dwarfs with
effective temperatures less than 50,000~K contain detectable metals
\citep{barstow03}.  DA white dwarfs hotter than 40,000~K can contain
elements heavier than H or He, such as  C, N, O, Si, P, S, Fe, and Ni in
their photospheres, giving rise to stellar absorption lines
\citep{venneslanz01,barstow03}.  Unlike rapidly rotating, early-type
stars that are typically used as background sources to study
interstellar absorption lines, the photospheric lines of white dwarfs
are not significantly rotationally broadened and can be easily confused
with interstellar or circumstellar features, especially if the radial
velocity of the star is unknown or is the same as the interstellar
velocity.   Nevertheless, white dwarf spectra generally have fairly
smooth continua making them excellent background sources for studying
the interstellar medium.

The 29 white dwarfs observed in this survey of the LISM are listed in
Table 1, along with their distances, \HI\ column
densities, Galactic coordinates, spectral
types, effective temperatures, and {\it FUSE} dataset names.  The
positions of the survey stars in Table 1 are plotted in Figures
1 and 2, projected onto the Galactic and Meridional planes,
respectively.  In these figures, we have plotted only the 25 stars
without strong stellar \OVI\ contamination, as described in Section 2.2
below. In Figure 1, only the subset of these 25 stars with $|b|<35^\circ$ are
displayed, to avoid misleading projection effects.  Overlayed in Figures
1 and 2 are the main boundaries of the 
20 m\AA\ \NaI\ D2 equivalent width contours from
\citet{lallement03}, which define the Local Cavity. A Na D2 line equivalent
width of 20 m\AA\ corresponds to an \HI\ column density of 
$\log N$(\HI\/)$=19.3$.
This contour level marks the beginning of a steep rise in density;
essentially defining an \HI\ ``wall'' surrounding
the local cavity.  Also shown in both
figures is the contour map of the hot LB, as defined by the 0.25 keV
soft x-ray emission \citep{snowden98}.

Considerable uncertainty exists in the distances to the white dwarfs and
in the location of the hot LB contour from the soft x-ray emission. 
Only two stars in our survey have parallaxes listed in the Hipparcos 
catalog \citep{perryman97} -- WD~0501$+$527 and WD~1314$+$293.  However,
even in these cases, the uncertainties are not small -- the
parallax error is 21\% for WD~0501$+$527 and 27\% for WD~1314$+$293.  Even more
worrisome is the fact that the Hipparcos-determined distance of 32 pc to
WD~1314$+$293 disagrees with numerous other determinations that all indicate a
distance of about 65 pc. \citet{kruk02} have discussed this in detail,
and we have adopted their assumed distance of $68\pm13$ pc.  The
distances to the remaining stars in Table 1 are taken from
\citet{vennes97} and \citet{holberg98}, who employ photometric techniques to
derive distances.  The star's spectral energy distribution is used
to determine the effective temperature and gravity of the star, and
these atmospheric parameters are used in conjunction with theoretical 
evolutionary sequences to determine the star's radius, and hence its
absolute magnitude. The distance is then determined from the star's apparent
magnitude.  Even though white dwarfs have well understood spectra, the
resultant uncertainty in distance is still $\sim 25-30$\%.

There is also significant uncertainty in the radial extent of the hot LB
as determined from soft x-ray emission.  In order to determine the
scaling relation between soft x-ray emission and distance,
\citep{snowden98} used a distance of $65\pm5$ pc to the molecular cloud
complex MBM 12 as determined from ISM absorption line measurements
\citep{hobbs86} together with X-ray shadowing measurements
\citep{snowden93}.  This distance determination is a matter of some dispute,
and the true distance may be considerably larger than this \citep{luhman01,
andersson02}.  We conclude that the LB may extend all the way
out to the boundary of the cavity demarked by the \NaI\ contours.

\citet{lehner03} points out that in the face of distance uncertainties,
it is also useful to consider the \HI\ column densities along the
lines of sight to the target stars.  Objects with column densities 
$\la 19.0$ dex are within the LB, while objects with $19.0 \la$ 
logN(\HI\/) $\la 19.3$
are probably very near the bubble wall, and stars with logN(\HI\/) $> 19.3$
are beyond the LB wall.  In our sample of 25 white dwarfs, 8 objects
have logN(\HI\/) $\ga 19.0$ (see Table 1).

\subsection{{\it FUSE} Data}

\citet{moos00} and \citet{sahnow00} have provided an overview of the
{\it FUSE} spectrograph and its on-orbit performance.  In
brief, the {\it FUSE} instrument consists of two telescopes, each of which has
two channels optimized for specific wavelength ranges. The SiC channels
cover the wavelength range 905-1104 \AA\/, while the LiF channels cover
the 990-1185 \AA\ region.  Hence, the \OVI\ \lam1032 line is covered by
both the SiC and LiF channels on sides 1 and 2 of the instrument.  Spectra 
of \OVI\ are obtained in
channels LiF1, LiF2, SiC1 and SiC2.  The SiC channels have only about
one-third the sensitivity of the LiF1 channel, and the LiF2 channel is
$\sim 60$\% as sensitive as LiF1 at 1032 \AA\/.  Since the \OVI\ line is
weak, the best data is usually  obtained in the LiF1 channel. 
Furthermore, in many of the observations obtained early in the {\it FUSE}
mission, the channels were not properly aligned, and hence for some
datasets, the target was not in the aperture of the SiC channels.  The
target was always in the LiF1  channel, since this channel contains the
guide camera.  Neverthless, we have used data from all channels if the
signal-to-noise is adequate. 

Some of the observations were taken in the MDRS (4 arcsec) and HIRS (1.5
arcsec) spectrograph apertures, although most of the data were obtained
in the LWRS (30 arcsec) aperture.  A typical observation consisted of
$10-20$ exposures, each of which was $\sim 10$ minutes in duration,
yielding total integration times of $5-10$ ksec per target. All data for
these stars were reprocessed with version 1.8.7 of the CALFUSE data
reduction pipeline.

The spectra produced by the {\it FUSE} concave gratings are astigmatic, with
significant curvature present in the two-dimensional point spread
functions.  However, the \OVI\ lines lie near one of two
astigmatic correction points, where the curvature cannot be easily
removed.  Hence, spectral extractions were performed by simply
collapsing the two-dimensional spectra to one-dimension. The resulting
spectral resolution is $\sim 15,000$ (FWHM) with no appreciable difference
between the MDRS and LWRS aperture data (the point spread function is
much smaller than the width of either aperture).

There are two sources of spectral motion in the {\it FUSE} instrument that can
degrade spectral resolution:  motions of the primary mirrors, and
rotation of the gratings in each  spectrograph channel.  Shifts
during an exposure are typically only $0.01$ \AA\/.   Consequently, prior
to coadding all the exposures in a given observation,  the spectra were
registered using the \CII\ \lam1036 line as a reference. 

At the present time, no correction for fixed pattern noise in the
detectors is available.  A typical resolution element at 1032 \AA\/
covers $\sim 100$ detector pixels (10 pixels in the dispersion direction
and 10 pixels of astigmatic height), which smooths out some of the 
pixel-to-pixel flat field irregularities.  For those spectra whose
exposures had some movement, the shift-and-add coaddition
process serves to further wash out some of the fixed pattern noise in
the spectra from each channel.  Whenever possible, we combined spectra
for a given star from all channels and all observations with weights 
inversely proportional to the noise variances.
Since the different channels are recorded on
different regions of the detector, this further mitigated the effect of
fixed pattern noise. 
Beginning in the second year of {\it FUSE} observations, we began to take data
in a special mode in an attempt to mitigate fixed pattern noise.  When
observing in the LWRS aperture, the target was observed at several
different locations within $\pm 10$ arcsec of the aperture center, resulting
in spectra at different locations on the detector.  Many, but not
all, of the observations with program ID P204 (see Table 1) were
obtained in this manner. 

Typical signal-to-noise ratios in the final coadded
spectra are $\sim 20$ to 50 per resolution element at 1032 \AA\/.  
Nevertheless, some fixed pattern noise remains in the data, and this
serves to limit our ability to measure weak absorption features.   

\citet{barstow01,barstow02} and \citet{wolff01} have reported
detection of stellar \OVI\ lines in the FUSE spectra of the hot 
(T$_{\rm{eff}} \ga$ 60,000 K) white dwarfs PG~1342$+$444,
RE~J0558$-$371 and Feige 55.
One surprising discovery from this survey was the existence of strong
photospheric \OVI\ lines in the spectra of two white dwarfs with
effective temperatures below 50,000 K:  WD~0131$-$163 and WD~2156$-$546 (see
Fig 3).  Stellar \OVI\ lines are also evident in the spectra of the
white dwarf in the close binary WD~2013$+$400 and the hot helium-rich
white dwarf WD~0501$-$289.  These stars all display a strong \OVI\ 1038 \AA\
line, with a 1032/1038 doublet ratio of approximately unity,
indicating that the lines are optically thick.  The photospheric
nature of the strong \OVI\ lines in these stars is confirmed by
comparing the \OVI\ radial velocities to those of other photospheric
lines.  We have eliminated these stars from further consideration in
this survey, since an accurate measurement of interstellar \OVI\ is not
possible.

There are several other features to note in Figure 3 that are the same
for all the white dwarf spectra.  First, interstellar  \CII\ \lam1036
and \OI\ \lam1039 absorption lines are present.  In most of the spectra
in our survey, the \CII\ \lam1036 line is lightly saturated with a full 
width half maximum (FWHM) of $\sim 100$ m\AA\/. In low column density
lines of sight, such as HZ 43, the \CII\ \lam1036  and \OI\ \lam1039
FWHM are $\sim 17$ \kms\/, representing the limiting resolution of the  
instrument at these wavelengths.  \CIIstar\ \lam1037 interstellar
absorption is also present if the electron density is high enough
somewhere along the line of sight.  Finally, we note the absence
of H$_2$ absorption,  which is a ubiquitous feature in far-UV spectra
along more distant Galactic sightlines, in all but 3 of our survey stars.
\citet{lehner03} reports total column densities of $\log$ N(H$_2$) 
$\sim 14.5-15$ towards WD~0004$+$330, WD~1636$+$351, and WD~1800$+$685.
Simply assuming an HD/H$_2$ ratio equal to the observed local D/H ratio of
$1.5 \times 10^{-5}$ \citep{linsky98, moos02}, 
would then imply an HD column density of
$\log N({\rm HD}) < 1.5 \times 10^{10}$ cm$^{-2}$.  However, \citet{watson73}
has pointed out that ion molecule, isotope exchange reactions in the ISM can
form HD preferentially in comparison to H$_2$, with enhancements of
order of $\sim 100$.  Even if one assumes such a large enhancement of HD,
we predict that the resulting equivalent width of the
6-0 R(0) Lyman transition of HD at 1031.909 \AA\/ would be $<0.25$ m\AA\/.
Hence, this HD line is not expected to contaminate the
\lam1032 line of \OVI\/ in the LISM.

\subsection{\OVI\ Measurements}

The zero point of the wavelength scale for calibrated {\it FUSE} spectra
taken in the LWRS aperture is uncertain by $\sim 15$ \kms\/, due to
uncertainty in the location of the star in the aperture, arising from
relative uncertainties of celestial coordinates of the target stars with
respect to guide stars (especially due to proper motion). Consequently,
we have used the \CII\ \lam1036.336 line to set the zero point for the
interstellar cloud velocity along each line of sight.  An additional
check of the \CII\ \lam1036 line was provided by the interstellar \OI\
\lam1039.23 line. A number of the white dwarfs in our survey also
displayed photospheric \PIV\ absorption lines at 1030.515 and 1035.516
\AA\/, which were used in conjunction with published stellar velocities
\citep{holberg98} to check the dispersion solution in the region
1030-1039 \AA\/.  Based on these interstellar and stellar lines, we are
confident that our corrected wavelength scale at 1032 \AA\ is accurate
to $< 10$ \kms\/. Spectra of the 25 white dwarfs, after excluding the
ones noted above as having strong stellar \OVI\ contamination, are shown
in Figure 4. These spectra have been normalized to the continuum, and
only the region near the \OVI\ \lam1032 line is shown. {\em The} \OVI\
{\em absorption line is remarkably weak in all sightlines.} Since the
\OVI\ \lam1031.93 line has an {\it f}-value  twice as large as the
1037.62 \AA\ member of the doublet, we have measured only the 1032 \AA\
line equivalent widths. After some experimentation, a velocity
integration range of $\pm40$ \kms\ was chosen in order to be large
enough to  fully cover the expected thermal width of an \OVI\ line at
$T=200,000$~K, plus some margin for wavelength calibration uncertainties
and real velocity offsets, but also narrow enough to exclude nearby
stellar features.  

The measured equivalent widths and column densities are given in Table
2. The column densities are derived from the measured equivalent
widths,  assuming a linear curve of growth.  The uncertainties in the
measured equivalent widths were determined empirically from the
following considerations.  In addition to fixed pattern noise, there is
also the likelihood that weak photospheric absorption lines from various
metals could be present in the spectra.   We have decided to treat these
lines as just another source of noise. Hence, we measured the equivalent
width in numerous regions of the spectra near the \OVI\ line where we
did not expect to find any stellar or interstellar lines; i.e., where the
expected equivalent width would be zero.   The standard deviation in the
measured equivalent widths was then used as our empirically determined
uncertainty. Consequently, the uncertainties in the equivalent widths in
Table 2 combine systematic and random errors.

No \OVI\ absorption was detected with significance greater than 2
$\sigma$ for 12 of the 25 stars. As can be seen from Table 2, a typical
1 $\sigma$ uncertainty is $\sim 4$ m\AA\ (equivalent to an \OVI\ column
density of $\sim 3 \times 10^{12}$ cm$^{-2}$).  Of the remaining stars,
most have column densities N(\OVI\/)$ < 10^{13}$ cm$^{-2}$ and no
column densities exceed $1.7 \times 10^{13}$ cm$^{-2}$.

\subsection{Photospheric and Circumstellar Contamination}

We now return to the issue of photospheric contamination of the \OVI\
interstellar lines.  As mentioned above, \OVI\/ lines can be formed in
the atmospheres of hot white dwarfs, even if they have  relatively low
metallicity. It was easy to reject the obvious cases where the
photospheric \OVI\ was strong and present in both members of the \OVI\
1032/1038 doublet.  But, what, might we ask, is the potential for
contamination by weak stellar \OVI\ lines?  
We have computed NLTE photospheric models of white dwarfs for a grid
of effective temperatures and atmospheric abundances assuming $\log g =   
8$ and a homogeneous H$+$O chemical composition.  The results are
illustrated in Figure 5.
The models indicate that stars with T$_{\rm{eff}} < 40,000$~K should not
have any appreciable absorption by \OVI\/.  Hence, the cool stars
WD~0050$-$332, WD~0549$+$158, WD~1017$-$138, WD~1254$+$223, WD~1615$-$154, 
WD~1636$+$351, 
WD~1845$+$019, WD~1847$-$223, and WD~2111$+$498 should not 
show photospheric \OVI\
lines.  Above this temperature, stars with low abundances can have weak
stellar absorption lines with strengths of a few m\AA\ rising to $\sim
10$ m\AA\ for T$_{\rm{eff}} = 60,000$~K and log(O/H)$<-7$.  
In Figure 6, we show the data for 5
of our survey stars, overlayed with photospheric models.   In these
models, the abundances of O, P and Fe were adjusted to give the best
fits to the data.  In the cases of P and Fe, there were multiple lines
with different strengths available for the fit; however, in the case of
oxygen, there is only the 1032/1038 doublet.  Typically, the best fitting
models have log(O/H)$ \approx -8$. In the case where the
stellar radial velocity and the interstellar velocity are quite
different, there is no ambiguity as to whether a line is stellar or
interstellar. For instance, as shown in the top panel of Figure 6, the
stellar \OVI\ line in WD~0455$-$282 is distinctly offset from the broad,
presumably interstellar \OVI\ line. However, in cases where the stellar
radial velocity is unknown (which, sadly, is the majority of the cases),
there is a general uncertainty as to the level of contamination of the
observed \OVI\ line by the star's photosphere for stars hotter than
40,000~K.   

Consequently, for the hotter stars, our \OVI\  measurements can be
viewed as providing upper limits to the interstellar absorption, since
some photospheric contribution may be present depending on the abundance
of oxygen in the star. For the models shown in Figure 6, the equivalent
width of \OVI\ reported for WD~0455$-$282 appears to not be
contaminated.   Also, for WD~2111$+$498, the \FeIII\ \lam1032.12 line
does not affect our measurement of \OVI\/, since it is outside our 40
\kms\ measurement window.  Note also that this star has an effective
temperature well below 40,000~K, so no  photospheric \OVI\ is predicted.
For the other three stars in Figure 6, it is possible to produce
photospheric models that can explain a significant part of the observed
\OVI\ absorption. We find that after subtracting our best fitting
photospheric model, that the interstellar \OVI\  equivalent widths for
WD~0501$+$527, WD~2211$-$495 and WD~2331$-$475 are 1.3, 5.6 and 5.4
m\AA\/, significantly lower than the measured values 
of 4.4, 16.4, and 13.1 m\AA\/, respectively.   These
model-dependent corrected values are noted in the footnotes to Table 2. 
At this point there is no satisfying model for explaining why the
photospheric \OVI\ abundances that we have measured are much smaller than
the \OIV\ and \OV\ abundances measured by \citet{venneslanz01} and 
\citet{barstow03}.
\citet{venneslanz01} suggested that this difference may be the
result of a steep oxygen abundance gradient that increases with
decreasing optical depths.  Consequently, we have assumed throughout the
rest of this paper that the measured column densities toward these stars
are formed in the ISM.

The \OVI\ line velocities reported in Table 2 are measured with respect
to the \CII\ \lam1036 line, which we use as the low ionization ISM cloud
velocity.  In this frame of reference, the \OVI\ line velocity is $\la
20$ \kms\ for most lines of sight.  The \OVI\ absorption along the line
of sight to WD0455$-$282, at $\sim -35$ \kms\/, is a notable exception.
The line of sight to this star appears more complicated than others in
the LISM.  The \CII\ \lam1036 interstellar line has multiple components;
the strong central one at 0 \kms\ (by definition) and another reasonably
strong line at $-54$ \kms\/.  The stellar lines in the {\it FUSE}
spectrum are at $\sim +60$ \kms\ relative to the central \CII\ ISM
line, in good agreement with the velocities of stellar and interstellar
lines reported by \citet{holberg98} from IUE observations. 
\citet{holberg98} also report the presence of  \SiIV\ and \CIV\
absorption lines  blueshifted  by $-53$ \kms\ with respect to the
stellar photosphere, which they cite as evidence for mass loss from this
hot, DA white dwarf.  We do not observe \OVI\ at this velocity.  We will
thus assume that the \OVI\ line, which appears to have multiple
unresolved components, is interstellar in nature.

\subsection{Line Widths}
  
In order to quantify the widths of the \OVI\ absorption lines, we have
computed their mean absolute deviations in order to avoid
the effects of noise in the wings of the line profile. The average
absolute deviation about the mean is defined as 
$v_{mad}={\sum_{i}f_i|v_i - \overline{v}|}/\sum_{i}f_i$, where $f_i$ is
the residual intensity and $\overline{v}$ is the line velocity centroid.
The results are
given in Table 2 for those lines of sight having sufficient \OVI\ line
strengths  and signal-to-noise to permit a measurement.  The average
value of $v_{mad}$ is $16\pm2$ \kms\/.  For a normal distribution, the
standard deviation is $1.25v_{mad}$, or 20 \kms\/. This observed line
width should be compared with the expected width, $\sigma_{tot}$,  of an
oxygen line formed in a $T=3 \times 10^5$~K gas, having a thermal width
of $\sigma_t = \sqrt{kT/16m_p}$ convolved with the {\it FUSE}  line spread
function, $\sigma_{lsf} \sim 8.5$ \kms\/.  The expected line width is
then   $\sigma_{tot}=\sqrt{\sigma_t^2 + \sigma_{lsf}^2}=15$ \kms\/,
which is less than our observed linewidths in the LISM.  This should be
compared  with the results of \citet{jenkins78b}, who found 
$\sigma_{tot} \sim 26$ \kms\ over much  longer pathlengths in the
Galactic disk.

\section{Discussion}

The weakness of \OVI\ absorption, not to mention its total absence along
many sightlines, in the LISM has important implications for models
of the Local Bubble, as well as our overall understanding of 
the interstellar medium.

\subsection{The Conductive Interface Model}

First, let us consider the implications of the observed results for the
conductive interface model.  \citet{slavin89} has modeled the
evaporation of the LIC inside the LB, assuming a temperature of $10^6$~K
and density of $n = 0.005$ cm$^{-3}$ in the LB, as implied by the soft
X-ray results.  The model includes the effects of radiative cooling,
non-equilibrium ionization, and magnetic fields.  Column densities of
\OVI\ through one cloud interface were computed as a function of the
parameter $\eta = {\rm cos}^2 \theta = 1 - B_T^2/B^2$, where $B_T$ is
the  tangential component of the magnetic field, and $\theta$ is the
angle of the magnetic field relative to radial.  If the magnetic field
in a cloud is largely tangential, then electrons, which travel along
field lines, are ineffective in transporting heat across the
interface.  This ``magnetic shielding'' prevents evaporation of the
cool cloud, and results in low \OVI\ column densities.  The column
densities predicted by Slavin's model for one interface range from $6.6
\times 10^{12}$ to $1.39 \times 10^{13}$ cm$^{-2}$ for $\eta = 0.1$ to
$0.9$, respectively.  About half of the stars in our survey within the
LB have \OVI\ column densities less than the model prediction even
assuming significant inhibition of thermal conduction by magnetic
fields (the $\eta=0.1$ case) for a {\it single} cloud interface.  Since
we expect that our lines of sight will intersect at least one cloud
interface (the LIC) and quite possibly one or more cloud
interfaces, the observations place severe constraints on the
effectiveness of thermal conduction in forming \OVI\ in cloud
interfaces.  Assuming that an ordered magnetic field is inhibiting conduction,
then the field vector must be $\ga 85^{\circ}$\ to the normal 
in order for \OVI\ column densities to be
as low as $\sim 10^{12}$ cm$^{-2}$.  A tangled magnetic field would also
be very efficient at halting conduction.
If the magnetic field becomes more radial over a
small region of the cloud surface and connects to the magnetic field in
the hot, intercloud medium, then conduction across the interface would
occur and create ``patches'' of \OVI\ on the cloud surface.  A line of
sight through the LB would encounter little \OVI\ absorption unless it
passed through one of these \OVI\ ``patches'' on a cloud surface. This
picture seems qualitatively consistent with the observations.

\citet{borkowski90} have computed models of plane-parallel magnetized
thermal conduction fronts and followed the evolution of the front to
predict column densities of selected UV absorption lines as a function
of time. Based on these models (see their Figure 6), our detections of
\OVI\ along some LISM sightlines with column densities exceeding $10^{12}
$ cm$^{-2}$ are consistent
with a lifetime of $\ga 10^5$ years for the Local Bubble. 
It could be argued then that this result does not
support the idea that the Local Bubble was re-heated and
re-ionized by a recent ($\la 10^5$ yrs) supernova explosion \citep{cr87}.
Alternatively, the overall weakness of
\OVI\ absorption along many sight lines would be consistent with recent
reheating, with detectable \OVI\ absorption being due to lines
of sight crossing multiple interfaces.

In Figure 7, we have plotted the column densities of \OVI\ versus \HI\/.
There is no obvious dependence. The \HI\ column is proportional
to the path length through cool clouds, and would therefore be expected
to correlate at least weakly with the number of cloud interfaces
intercepted along the line of sight.  The lack of a correlation 
between \OVI\ and \HI\ column density is consistent with the ``patchy''
\OVI\ interface model.

Understanding pressure balance within the LB has been a long-standing
problem.  In the LIC, the observed pressure is $P/k = 2nT \sim 2000$
cm$^{-3}$~K \citep{redfield00, jenkins02},  while in the hot, intercloud medium the
value is $\sim 10,000$ cm$^{-3}$~K based on temperatures and densities
derived from the soft X-ray data.  By invoking a magnetic field of
$5 \mu$G, a magnetic pressure of $(P_B/k) \sim 7500$ cm$^{-3}$~K provides
the additional pressure needed to keep the cool clouds in the LB from
being crushed. The unexpectedly weak \OVI\ absorption observed in the
LISM, if such weakness is attributed to the quenching of conduction, 
provides some evidence for the existence of magnetic fields within the
local clouds, although we cannot derive an absolute measure of the
magnetic pressure.  The severity of this pressure balance problem is 
ameliorated if the observed soft x-ray emission is significantly
contaminated by a local heliospheric contribution \citep{robertson01}. 
A lower x-ray flux from the LB would translate directly to a lower
density, and hence a lower value of $P/k$ in the hot medium.  

Our observations are also consistent with the recent
model for the origin of cool clouds inside the LB by \citet{ch03}. 
These authors propose that magnetic flux tubes separate from the inner
wall of the bubble except at their anchored ends, and  through their
tension, spring into the interior of the bubble.  In doing so, the flux
tubes drag material into the cavity.  In this scenario, the magnetic
field is not tangled, but is ordered parallel to the cloud surfaces, and
will be effective in preventing conduction over a substantial
portion of the surface.

\citet{vallerga98} has shown that the spectrum of the
EUV flux from local stellar sources (B stars and white dwarfs) is
too soft to explain the overionization of helium with respect to hydrogen 
in the LISM \citep{dupuis95}.  As discussed earlier, \citet{slavin02}
have shown that radiation from an evaporative interface is required
in order to correctly predict the level of ionization in the LISM.
The observed weak \OVI\/ absorption, and the
interpretation that this is due to the inhibition of thermal conduction, 
then leaves the He ionization problem without an obvious solution.

\subsection{\OVI\ Sky Distribution and Space Density}

In Figure 8, we show the distribution of \OVI\ absorption on the
sky.  In this plot, the size of the circle around each survey star is
inversely proportional to its distance.  Each circle has a gray-scale
shading proportional to the column density along that line of sight,
with darker shades representing higher column densities. The lines of
sight with Galactic longitudes $210^\circ \la l \la 340^\circ$ and
negative latitudes appear to have higher \OVI\ column densities than 
elsewhere, although this trend is somewhat weakened by the likelihood
that the \OVI\ column densities towards WD~2211$-$495
and WD~2331$-$475,  which are both at approximately Galactic coordinates
($l \sim 340^\circ$, $b \sim -60^\circ$), are contaminated by
photospheric absorption (see \S 2.4).

Armed with column densities and distances, we can compute the space
density of \OVI\/, $\langle n($\OVI\/$) \rangle$ (in cm$^{-3}$). There
are several different methods that can be employed to do this.
\citet{jenkins78b} computed the value of $\Sigma
N/\Sigma r$; the total column density of \OVI\ for all lines of sight
divided by the total effective path length to all stars (the effective
path length is the equivalent distance of the star if it were in the
Galactic plane, taking into account the decrease in density of \OVI\
away from the plane).  Using this definition, \citet{jenkins78b} found
an \OVI\ density in the Galactic plane of $2.8 \times 10^{-8}$ cm$^{-3}$, 
or $2.1 \times 10^{-8}$ cm$^{-3}$ after discarding several lines of
sight with ``anomalous'' column densities.  Another method is just to
compute the mean of all values of $N_i($\OVI\/$)/r_i$.  
However, this method tends
to overweight lines of sight with anomalous densities. 
This can be alleviated by using an inverse variance weighting,
$$\langle n(\mathrm{OVI}) \rangle = s^2{\sum_{i} {N_i \over r_i}\sigma_i^{-2}}$$
where $s^2 = 1/{\sum_{i} \sigma^{-2}}$ is the variance of 
$\langle n($\OVI\/$) \rangle$, $\sigma_i$ is the uncertainty
in $(N_i/r_i)$, $N_i$ is
the column density of \OVI\ and $r_i$ the effective distance,
respectively, for the $i$th star.  Using all 25 stars in the sample, we
find an inverse variance weighted mean density of $1.9 \pm 0.25 \times
10^{-8}$ cm$^{-3}$, and an unweighted mean density is $2.45 \times
10^{-8}$ cm$^{-3}$.  The value computed from  $\Sigma N/\Sigma r$ is
approximately the same at  $2.36 \times 10^{-8}$ cm$^{-3}$, and is
comparable to the value found by \citep{jenkins78b} over much larger
distances in the Galactic disk.  A more recent determination of the
Galactic disk \OVI\ density has been made from a FUSE survey of 150
stars
\citep{jenkins01, bowen04}.  They find a median value of $N_i($\OVI\/$)/r_i$
(measured for each sightline $i$)
of $1.7 \times 10^{-8}$ cm$^{-3}$.   We find a median value of the
\OVI\ density from our LISM survey of $1.9 \times 10^{-8}$ cm$^{-3}$.  
In summary, the space density of \OVI\ that we observe locally is about
the same, or somewhat larger than, the average for the Galactic disk
over kpc scales.  

\citet{sc94} reanalyzed the {\it Copernicus} data and included a contribution
to the \OVI\ column density from the Local Bubble along all lines
of sight.  Based on the Copernicus data alone, which had few stars
at distances of less than 100 pc, they concluded that the
LB contributed an \OVI\ column density of 
$1.6 \times 10^{13}$ cm$^{-2}$, and hence the mean space density
in the Galactic plane outside the LB was lower than found by 
\citep{jenkins78b}.  However, our survey indicates that the 
average \OVI\ column density out to 
a distance of 100 pc from the Sun
(the approximate distance to the edge of the LB in the Galactic plane)
is $\sim 7 \times  10^{12}$ cm$^{-2}$, based on our measured unweighted
mean space density of $2.45 \times 10^{-8}$ cm$^{-3}$.  

We have performed some numerical simulations which show that our
LISM survey is consistent with a scenario in which $\sim 6$ \OVI\
producing clouds or interfaces
per kpc are randomly distributed in space, each with a column density
of $\sim 1.25 \times 10^{13}$ cm$^{-2}$.  This distribution of clouds
predicts a dispersion in the \OVI\ space densities that is consistent
with that measured for
the 25 lines of sight in the LISM survey.  Hence, \OVI\
absorption in the LISM appears to have a distribution quite similar
to, or slightly larger than, much 
longer lines of sight in the Galactic disk.  We find this to
be a somewhat remarkable result, since we know that the Sun sits
roughly in the middle of a supernova cavity -- the LB.  If hot bubbles
only make up $\sim 20$\% by volume of the Galactic disk, then the 
distance between bubbles should be fairly large.  If \OVI\ is formed
in these hot regions of space, either at the interfaces with cool clouds or in
condensation regions inside the bubbles, then the volume density of \OVI\ 
producing regions in the disk should be lower than it is locally.
Possibly the discrepancy is due to the quenching of conduction by
magnetic fields on local cloud surfaces (i.e. without such quenching, the 
local space density of \OVI\ would be higher), and/or that the conditions
under which \OVI\ is formed locally is not representative of the
Galactic disk in general.

In the \citet{mo77} theory of the ISM, cool clouds reside in a
pervasive, hot ($T=10^6$~K) interstellar medium which is thermally
stabilized by cloud evaporation.  However, if conduction at cloud
interfaces is largely quenched, then there will be little or no cloud
evaporation. If magnetic shielding of conduction is common
throughout the disk (as
proposed here for the LISM), then the structure of the general ISM must
be quite unlike that proposed by \citet{mo77}.

\subsection{The Delayed Recombination Model}

\citet{breit94} have proposed a ``delayed recombination'' model 
for the Local Bubble in which an adiabatic expansion of hot gas is powered by
supernovae. The gas cools quickly as it expands, and highly ionized stages are
``frozen in'' to the gas.  The observed soft X-ray emission is then due 
to delayed recombination at low kinetic temperatures. In this model, the
intercloud  medium has a kinetic temperature of $\sim 50,000$~K and an
electron density of $0.024$ cm$^{-3}$, yielding $P/k \sim 2600$
cm$^{-3}$ K. This ameliorates the pressure balance problem, which is a
strong  selling point of this model.  Let us now consider the expected
\OVI\ production within the LB from this model.  The column density of
\OVI\ along a path of length $l$ is  $N($\OVI\/$) \approx \xi($\OVI\/$)
\delta($O$) X($O$)n_e(n_H/n_e)l$  cm$^{-2}$, where $\xi$(\OVI\/) is the
ionization fraction of \OVI\/, $\delta$(O) 
is the linear depletion of oxygen onto
dust grains, $X$(O) is the abundance of oxygen
relative to hydrogen, and $n_H/n_e = 0.83$ for a fully ionized gas that is 10\%
helium.  \citet{sm76} have computed models for the non-equilibrium
radiative cooling of an optically thin gas which has been shock heated
to $10^6$~K and allowed to cool to $10^4$~K, which is  less extreme than
the adiabatic cooling in the  \citet{breit94} model.  Even after the gas
has cooled to $T=5 \times 10^4$~K, there is still significant fractions
of O$^{+5}$ and O$^{+6}$ in the gas -- in particular, $\xi$(\OVI\/)$
\sim 0.015$ (see Figure $5d$ in \citet{sm76}). Adopting a solar
abundance of oxygen of $X($O$) = 4.57 \times 10^{-4}$
\citep{asplund04}, a path length of 100 pc, and  virtually no depletion
($\delta($O$)=1.0$) \citep{sofia01},  we then arrive at an expected
column density of \OVI\ of $N($\OVI\/$) \sim 4.2 \times 10^{13}$
cm$^{-2}$.  This column density is ruled out by the observations
reported here.

\subsection{Supernova Cavity Models}

\citet{coxsmith74}, \citet{slavin92}, \citet{shelton99} and \citet{sc01}
have investigated the formation of \OVI\ in the cavities carved out in
the ISM by supernova explosions.   In these models, the \OVI\ resides in
two distinct regions, the hot interior and the cooler periphery.  Each
has different characteristics and will be discussed separately. Here, we
will concentrate on the recent computations by \citet{sc01} of a
multiple explosion scenario for the formation of the LB.  Their models B
and G, in which 3 explosions occur during a 2 million year period, give
results that are in relatively good agreement with soft x-ray and \OVI\ 
absorption line results.  

In these models, after $\sim 6$ Myr, the
interior of the 100 pc radius cavity is still highly ionized. Assuming a
density of $n \sim 0.005$ cm$^{-3}$, $T=10^6$~K, an ionization fraction
of \OVI\ of $\xi$(\OVI\/) $=0.006$ \citep{sm76}, a path length of 75 pc,
and no  depletion, we expect an \OVI\ column density of  $N$(\OVI\/)$
= 2.6 \times 10^{12}$ cm$^{-2}$.  A column density of this level 
thermally broadening in a million degree gas
is too difficult to detect with the signal-to-noise in the current dataset.  

\OVII\ is the dominant ion in the outer parts of the cavity, where the
gas density is higher and recombination is more effective. \OVI\ is
abundant only near the edge of the cavity.  \citet{sc01} indicate that
the model most appropriate for the LB is intermediate between their
models B and G, for which they predict average column densities for
\OVI\ of 0.8 and $2.8 \times 10^{13}$ cm$^{-2}$, respectively, for a
line of sight from the center to edge of the bubble.  This estimate is
for the diffuse medium only, and does not include any contribution from 
interfaces of cool clouds embedded in the LB. 

Let us restrict our attention to the stars in our LISM survey for which
the Galactic latitude, $|b|<35^\circ$, so that the star's distance is
more closely related to the distance to the bubble wall (e.g. it can be
seen from Figure 2 that the LB has a ``chimney'' appearance, so that
stars at high Galactic  latitude are far from the wall of the local
cavity).  Thirteen stars in our survey satisfy this constraint on
Galactic latitude. In Figure 9, we have plotted \OVI\ column density
versus distance of the star from the Sun for these 13 stars, plus the
data for stars at distances of $100-400$ pc with $|b|<35^\circ$ from the
earlier {\it Copernicus} study by \citet{jenkins78a}. There does not
appear to be any convincing evidence for a sharp rise in \OVI\ column
density at a distance corresponding to the edge of the  cavity from this
plot.  In Figures 1 and 2 we also show the location of the stars in
the LISM survey relative to the cavity wall (as defined by
\citet{sfeir99}), and we find that there is no correlation between the
strength of \OVI\ absorption and closeness to the wall.  These
conclusions, however, are based on a relatively small sample of lines
of sight.  Finally, we also point out that the  \OVI\ column density
intermediate between models B and G \citep{sc01} of $\sim 2 \times
10^{13}$ cm$^{-2}$ is substantially larger than the observed mean
column density of  $\sim 7 \times 10^{12}$ cm$^{-2}$ within the LB
(assuming a path length of 100 pc).  The eight stars in our survey
having large \HI\ column densities, logN(\HI\/)$>19.0$, 
implying that they are near the wall of the cavity, have an average
\OVI\ column density of $5.8 \times
10^{12}$ cm$^{-2}$, which is also substantially less than the
column density predicted by models B and G.

\section{Conclusions}

We have carried out a survey of interstellar \OVI\ absorption along the
lines of sight to 25 white dwarfs in the LISM, after excluding 4 stars
that display strong photospheric \OVI\/.  \OVI\ absorption is weak or
absent in most lines of sight, and has a somewhat ``patchy''
distribution (a scatter that is much larger than the measurement
uncertainties).  The \OVI\ absorption in several stars may be due in
part to  photospheric or circumstellar absorption, which would imply
even smaller column densities in the ISM for these lines  of sight.  We
find a mean line width, based on computing mean absolute deviations for
the \OVI\ line profiles, of $16\pm2$ \kms\/, corresponding to a standard
deviation of 20 \kms\/.  This line width is about 30\% larger than
expected for a single component absorption line formed at 300,000~K and
convolved with the {\it FUSE} instrumental resolution. 

The general picture that emerges from these observations is one in which
the observed \OVI\ absorption is produced over limited regions on the
surfaces of clouds within the Local Bubble.  Conduction must be quenched
over much of the surface area of each cloud, presumably by magnetic
fields.  Indeed, we know that the Sun is located within  a cool cloud 
inside the hot LB, and all lines of sight should penetrate at least one
cloud interface, if not several.  Quite sophisticated theoretical models
on the other hand indicate that an unshielded interface should produce
an \OVI\ column density of $\sim 10^{13}$ cm$^{-3}$, yet many of our
lines of sight indicate columns significantly below this value.  The
mean \OVI\ space density in our LISM survey is $\sim 2.4 \times 10^{-8}$
cm$^{-3}$, and is comparable to, or slightly larger than, the value over
much larger  path lengths (400-1000 pc) in the Galactic plane determined
by the {\it Copernicus} survey \citep{jenkins78b}.   Out to a distance of
$\sim 100$ pc from the Sun,  corresponding roughly to the size of the
Local Bubble in the Galactic plane, the average column density of \OVI\
is $\sim 7 \times 10^{12}$ cm$^{-2}$, or about half that estimated
earlier \citep{sc94}.

This picture, however, implies that there should be little EUV radiation
from the evaporative interface between the LIC and the presumably hot
interior of the Local Bubble, at odds with the models of \citet{slavin02},
which have successfully explained a number of observations of
the ionization in the LISM.

\acknowledgments

This work is based on data obtained for the Guaranteed Time Team by the
NASA-CNES-CSA {\it FUSE} mission operated by the Johns Hopkins
University. Financial support to U. S. participants has been provided by
NASA contract NAS5-32985. We thank Jeff Kruk for informative discussions
of white dwarf spectra, John Vallerga for insightful comments on an
earlier version of the paper, 
and Barry Welsh for keeping us informed of his
studies of the LISM.  The authors would like to especially thank the
{\it FUSE} mission operations team at Johns Hopkins University, and in
particular, Alice Berman, Bill Blair, Humberto Calvani, Bryce Roberts
and Martin England, who made these observations possible through their
hard work and dedication.

\clearpage


\begin{deluxetable}{llccrrlrl}
\tablecolumns{9}
\tablewidth{0pt}
\tablecaption{Survey Star Data}
\tablehead{
\colhead{WD Number} &
\colhead{Alt. Name} &
\colhead{Distance\tablenotemark{a}} &
\colhead{logN(\HI\/)\tablenotemark{b}   } &
\colhead{$l$} &
\colhead{$b$} &
\colhead{Type\tablenotemark{a}} &
\colhead{T$_{\rm{eff}}$\tablenotemark{a}} &
\colhead{Datasets\tablenotemark{c}}  \\

\colhead{} & \colhead{} & \colhead{pc} & \colhead{cm$^{-2}$}  & \colhead{deg} & \colhead{deg} &
\colhead{} & \colhead{(K)} & \colhead{}
} 
\startdata
0004$+$330 & GD 2               & 108 & 19.80 & 111 & $-28$ & DA1 & 49,360 & P2041101 \\
0050$-$332 & GD 659             &  58 & 18.40 & 299 & $-84$ & DA1.5 & 36,000 & P2042001 \\
0131$-$163 & GD 984             & 104 & ...   & 167 & $-75$ & DA1+dM? & 48,700 & P2041201 \\
0455$-$282 & RE~J0457$-$281     & 102 & 18.06 & 222 & $-37$ & DA1 & 57,200 & P1041101 \\
           &                    &     &       &     &      &     &        & P1041102 \\
           &                    &     &       &     &      &     &        & P1041103 \\ 
0501$+$527 & G191$-$B2B         & 69  & 18.30 & 156 & $+7$  & DA1 & 56,000 & S3070101 \\
           &                    &     &       &     &      &     &        & P1041201 \\
           &                    &     &       &     &      &     &        & P1041202 \\
0501$-$289 & RE~J0503$-$289     & 90  & ...   & 230 & $-35$ & DO  & 70,000 & P2041601 \\
0549$+$158 & GD 71              & 49  & 17.92 & 192 & $-5$  & DA1.5 & 32,750 & P2041701 \\
0715$-$704 & ...                & 94  & 19.32 & 282 & $-23$ & DA  & 43,600 & P2042101 \\
1017$-$138 & ...                & 90  & 18.68 & 256 & $+35$ & DA  & 32,000 & P2041501 \\
1211$+$332 & HZ 21              & 115 & ...   & 175 & $+80$ & DO2 & 53,000 & P2040801 \\
           &                    &     &       &     &       &     &        & P2040802 \\
1234$+$481 & HS1234$+$481       & 129 & 19.13 & 130 & $+69$ & DA1 & 56,400 & P2040901 \\
1254$+$223 & GD 153             & 73  & 17.92 & 317 & $+84$ & DA1.5 & 38,686 & M1010401 \\ 
           &                    &     &       &     &       &       &        & M1010402 \\
           &                    &     &       &     &       &       &        & M1010403 \\
           &                    &     &       &     &       &       &        & P2041801 \\
1314$+$293 & HZ 43              & 68  & 17.94 & 54  & $+84$ & DA1   & 50,560 & P1042301 \\
1529$+$486 & RE~J1529$+$483     & 184 & 18.30 & 79  & $+53$ & DA1   & 47,600 & P2040101 \\
1615$-$154 & EG 118             & 55  & ...   & 359 & $+24$ & DA1.5 & 29,732 & P2041901 \\
1631$+$781 & RE~J1629$+$780     & 67  & 19.30 & 111 & $+33$ & DA1+dMe & 44,560 &P1042901 \\
           &                    &     &       &     &       &       &   & P1042902 \\
1634$-$573 & HD149499B          & 34  & 18.80 & 330 & $-7$  & DO+KOV & 49,500 & M1031103 \\
           &                    &     &       &     &       &   &  &    M1031105 \\ 
           &                    &     &       &     &       &   &  &    S5140201 \\
1636$+$351 & KUV 16366$+$3506   & 113 & 19.57 & 57  & $+41$ & DA   & 37,200 & P2040201 \\
1800$+$685 & KUV 1800$+$685     & 134 & 18.86 & 99  & $+30$ & DA1  & 46,000 & P2041001 \\
1845$+$019 & Lanning 18         &  44 & 18.45 & 34  & $+2$  & DA1.5 & 30,000 & P2040301 \\ 
1847$-$223 & ...                & 62  & 19.55 & 13  & $-9$  & DA   & 31,600  & P2040501 \\
2004$-$605 & RE~J2009$-$602     & 62  & 19.32 & 337 & $-33$ & DA1  & 44,200 & P2042201 \\
2013$+$400 & RE~J2013$+$400     & 143 & ...   & 78  & $+3$  & DAO+dM  & 53,600 & P2040401 \\
2111$+$498 & GD 394             & 50  & 18.60 & 91  & $+1$  & DA1.5 & 37,360 & M1010704 \\
           &                    &     &       &     &       &      &        & M1010706 \\
           &                    &     &       &     &       &      &        & P1043601 \\
2127$-$222 & ...                & 224 & 18.43 & 27  & $-44$ & DA   & 49,800 & P2040601 \\
2156$-$546 & RE~J2156$-$543     & 109 & ...   & 340 & $-48$ & DA   & 45,800 & P2042301 \\
2211$-$495 & RE~J2214$-$491     & 55  & 18.30 & 346 & $-53$ & DA1  & 63,500 & M1030305 \\
           &                    &     &       &     &       &      &        & P1043801 \\
2309$+$105 & GD 246             & 72  & 19.12 & 87  & $-45$ & DA1  & 58,700 & P2042401 \\
2331$-$475 & RE~J2334$-$471     & 82  & 17.90 & 335 & $-65$ & DA1  & 55,800 & P1044201 \\
\enddata
\tablenotetext{a}{Sources: \citet{holberg98, vennes97}  }
\tablenotetext{b}{Sources: \citet{dupuis96}, \citet{marsh97}, 
\citet{holberg98}, \citet{wolff99} }
\tablenotetext{c}{{\it FUSE} Observation dataset name, consisting of a 4-digit
program name, followed by a 2-digit target identifier and a 2-digit
observation number}
\end{deluxetable}


\begin{deluxetable}{lrrrrc}
\tablecolumns{6}
\tablewidth{0pt}
\tablecaption{\OVI\ \lam1032\ Measurements}
\tablehead{
\colhead{WD Number} &
\colhead{W$_\lambda$} &
\colhead{N(\OVI\/)} &
\colhead{${\bar v}$\tablenotemark{a}} &
\colhead{$v_{mad}$\tablenotemark{b}}  &
\colhead{Notes}  \\
\colhead{} & \colhead{(m\AA\/)} & \colhead{($10^{13}$ cm$^{-2}$)} &
\colhead{(\kms\/)} & \colhead{(\kms\/)} 
} 
\startdata

  0004$+$330   &   $  6.9\pm2.4$   &   $ 0.55\pm0.19$ & $+7$    &  ...  & \\
  0050$-$332   &   $ 11.3\pm4.1$   &   $ 0.90\pm0.33$ & $+20$   &  $18$ & \\
  0455$-$282   &   $ 16.6\pm3.7$   &   $ 1.33\pm0.30$ & $-35$   &  $18$ & 1 \\
  0501$+$527   &   $  4.4\pm2.8$   &   $ 0.35\pm0.22$ & $+5$    & ...   & 2 \\
  0549$+$158   &   $ -0.7\pm4.1$   &   $-0.06\pm0.33$ & ...     & ...   &  \\
  0715$-$704   &   $ 21.1\pm4.0$   &   $ 1.69\pm0.32$ & $-4$    &  $20$ & \\
  1017$-$138   &   $ 15.5\pm8.6$   &   $ 1.24\pm0.69$ & ...     & ...   & \\
  1211$+$332   &   $ -0.4\pm3.7$   &   $-0.03\pm0.30$ & ...     & ...   &  \\
  1234$+$481   &   $  5.3\pm8.3$   &   $ 0.43\pm0.66$ & ...     & ...   &  \\
  1254$+$223   &   $ 10.7\pm4.2$   &   $ 0.86\pm0.33$ & $+31$   &  $16$ & \\
  1314$+$293   &   $  8.7\pm1.9$   &   $ 0.70\pm0.15$ & $+20$   &  $15$ & \\
  1529$+$486   &   $ 18.8\pm3.5$   &   $ 1.50\pm0.28$ & $+4$    &  $17$ & \\
  1615$-$154   &   $ -0.1\pm6.0$   &   $ 0.00\pm0.48$ & ...     & ...   &  \\
  1631$+$781   &   $  3.7\pm3.3$   &   $ 0.30\pm0.26$ & ...     & ...   &  \\
  1634$-$573   &   $  7.8\pm3.4$   &   $ 0.63\pm0.27$ & $0$     &  $13$ & \\
  1636$+$351   &   $ 10.1\pm4.6$   &   $ 0.81\pm0.37$ & $+19$   &  ...  &  \\
  1800$+$685   &   $  9.6\pm5.7$   &   $ 0.77\pm0.45$ & $+12$   & ...   &  \\
  1845$+$019   &   $ -3.9\pm6.8$   &   $-0.31\pm0.55$ & ...     & ...   &  \\
  1847$-$223   &   $ -1.7\pm4.6$   &   $-0.13\pm0.37$ & ...     & ...   &  \\
  2004$-$605   &   $ 12.5\pm3.6$   &   $ 1.00\pm0.29$ & $+2$    &  $17$ & \\
  2111$+$498   &   $  2.1\pm5.6$   &   $ 0.17\pm0.45$ & ...     & ...   &  \\
  2127$-$222   &   $  9.6\pm4.4$   &   $ 0.77\pm0.35$ & $0$     & ...   & \\
  2211$-$495   &   $ 16.4\pm3.7$   &   $ 1.32\pm0.29$ & $+25$   &  $16$ & 2 \\
  2309$+$105   &   $  0.4\pm2.8$   &   $ 0.03\pm0.23$ & ...     & ...   &  \\
  2331$-$475   &   $ 13.1\pm3.7$   &   $ 1.05\pm0.30$ & $+19$   &  $15$ & 2 \\
\enddata
\tablecomments{Equivalent widths were measured in the
velocity interval $\pm40$ \kms\ centered on the expected position of \OVI\
\lam1036\ relative to the \CII\ \lam1036 line. 
(1) Significant absorption occurs outside the 
velocity integration limits for WD~0455$-$282. (2) The \OVI\ line measurement 
is likely contaminated by photospheric absorption; see \S 2.4 for discussion.}  
\tablenotetext{a}{Velocity of the \OVI\ \lam1032 line with respect to the 
ISM as defined by the \CII\ \lam1036 line.}
\tablenotetext{b}{\OVI\ \lam1032 line width computed from a mean absolute 
deviation. See \S 2.5 for discussion.} 
\end{deluxetable}

\clearpage

\begin{figure*}  
\plotone{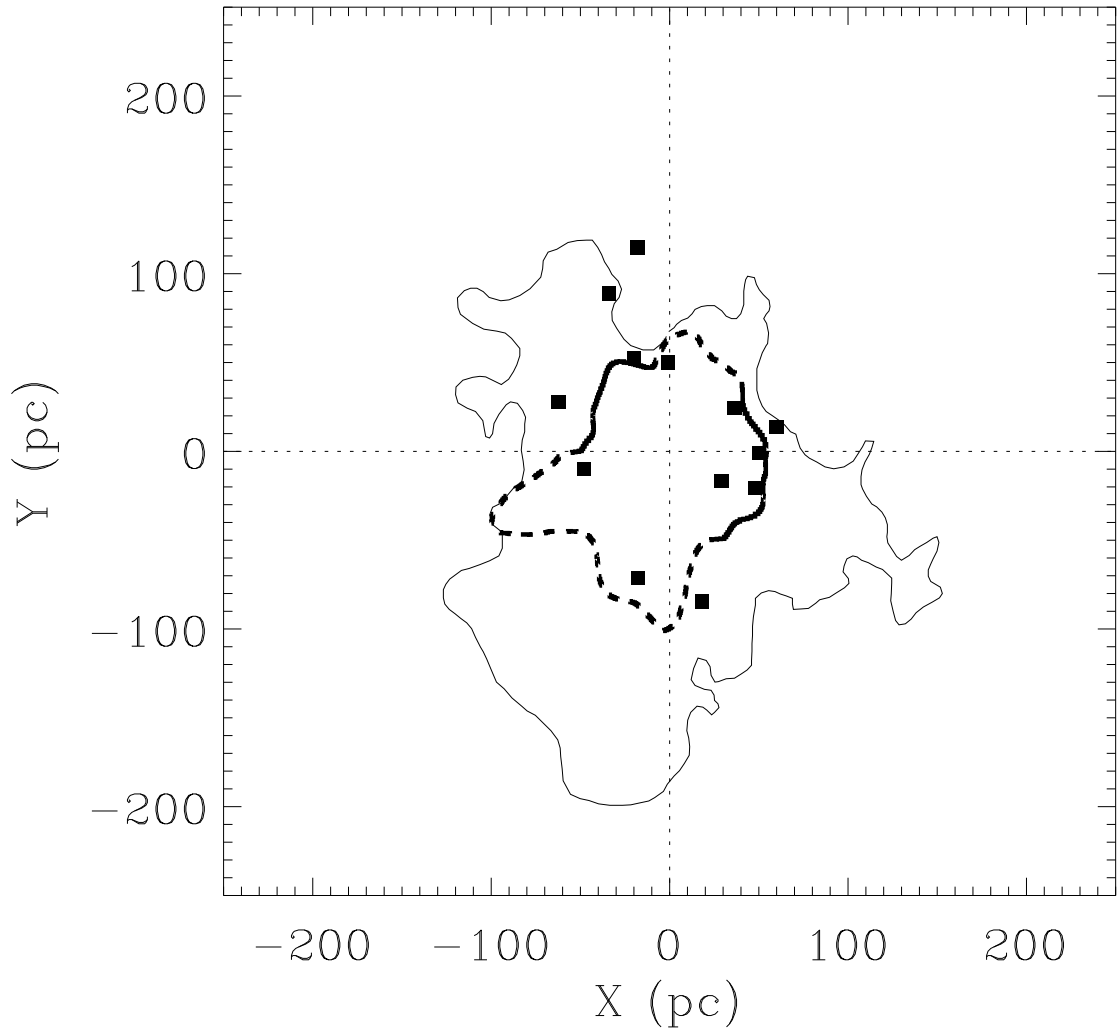}
\caption{Positions of the 13 white dwarfs from Table 1 with $|b|<35^\circ$, 
projected onto the
Galactic Plane (solid squares).  Overlayed on the plot (thin solid line) 
is the boundary
of the Local Cavity as defined by 20 m\AA\ Na D absorption line 
equivalent widths from \citet{lallement03}.  The thick solid line is 
the boundary of the hot Local Bubble, as defined by the 0.25 keV  
soft x-ray emission (regions affected by contaminating sources are
represented by dashed lines) \citep{snowden98}.  The distance to the
boundary of the LB is a matter of some debate (see text), and may
coincide with the Local Cavity. The Galactic center is in the
direction of the positive X axis, and Galactic longitude, $l=90^\circ$,
is in the positive Y axis direction.}
\end{figure*}

\begin{figure*}
\plotone{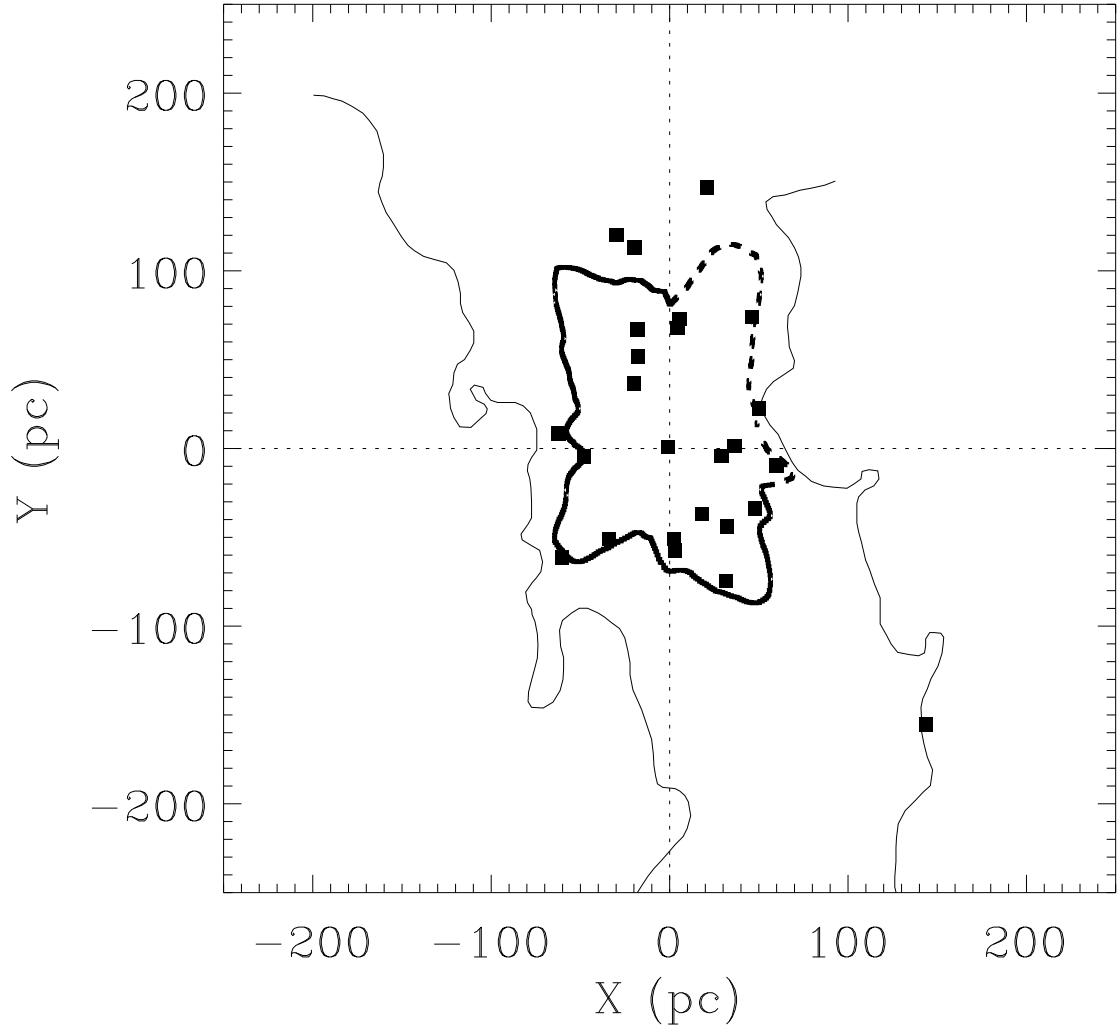}
\caption{Same as Figure 1, except this is a projection onto 
the Meridional plane, for all 25 white dwarfs listed in Table 1.
The Galactic center is in the direction of the positive X axis, 
and Galactic latitude, $b=90^\circ$, is in the positive Y axis direction.}
\end{figure*}

\begin{figure*}
\plotone{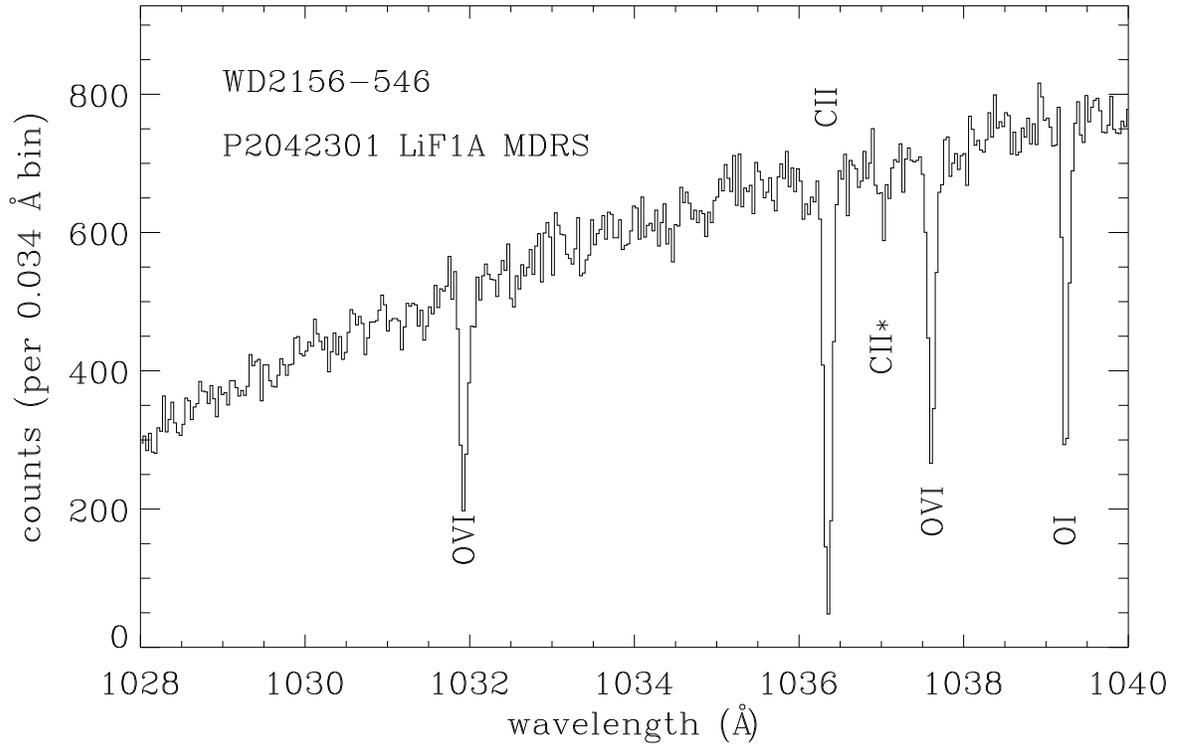}
\caption{Spectrum of WD~2156$-$546 taken through the MDRS aperture in the
LiF1A channel, showing strong photospheric absorption in both members of
the \OVI\ doublet at \lam1031.93 and \lam1037.62.  Also present in the
spectrum are interstellar lines of \CII\ \lam1036.34 and \OI\ \lam1039.23.}
\end{figure*} 

\begin{figure*}
\epsscale{0.7}
\plotone{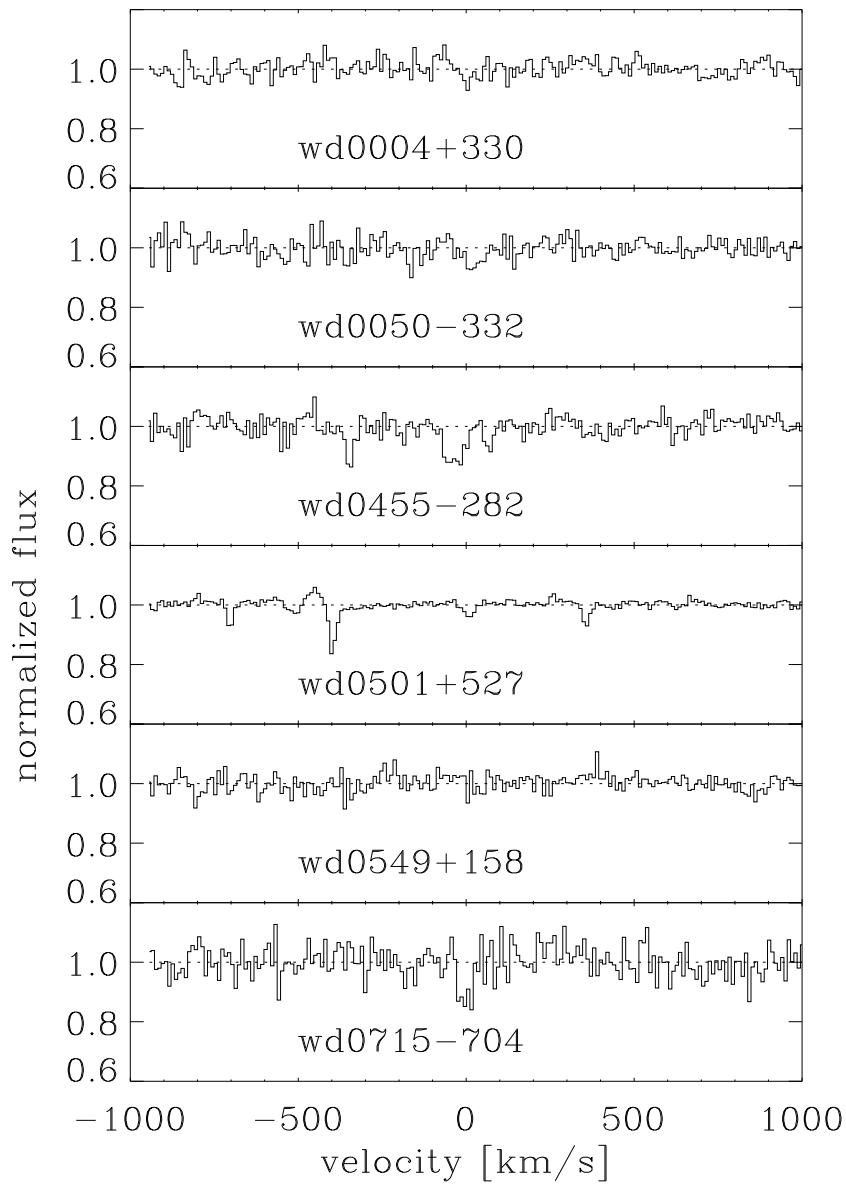}
\caption{Normalized spectra of the 25 white dwarfs from Table 2
are plotted, centered on
the \OVI\ \lam1031.926 line as a function of velocity.
The velocity zero point
is set relative to the centroid of the \CII\ \lam1036.336 line.}
\end{figure*}

\begin{figure*} 
\plotone{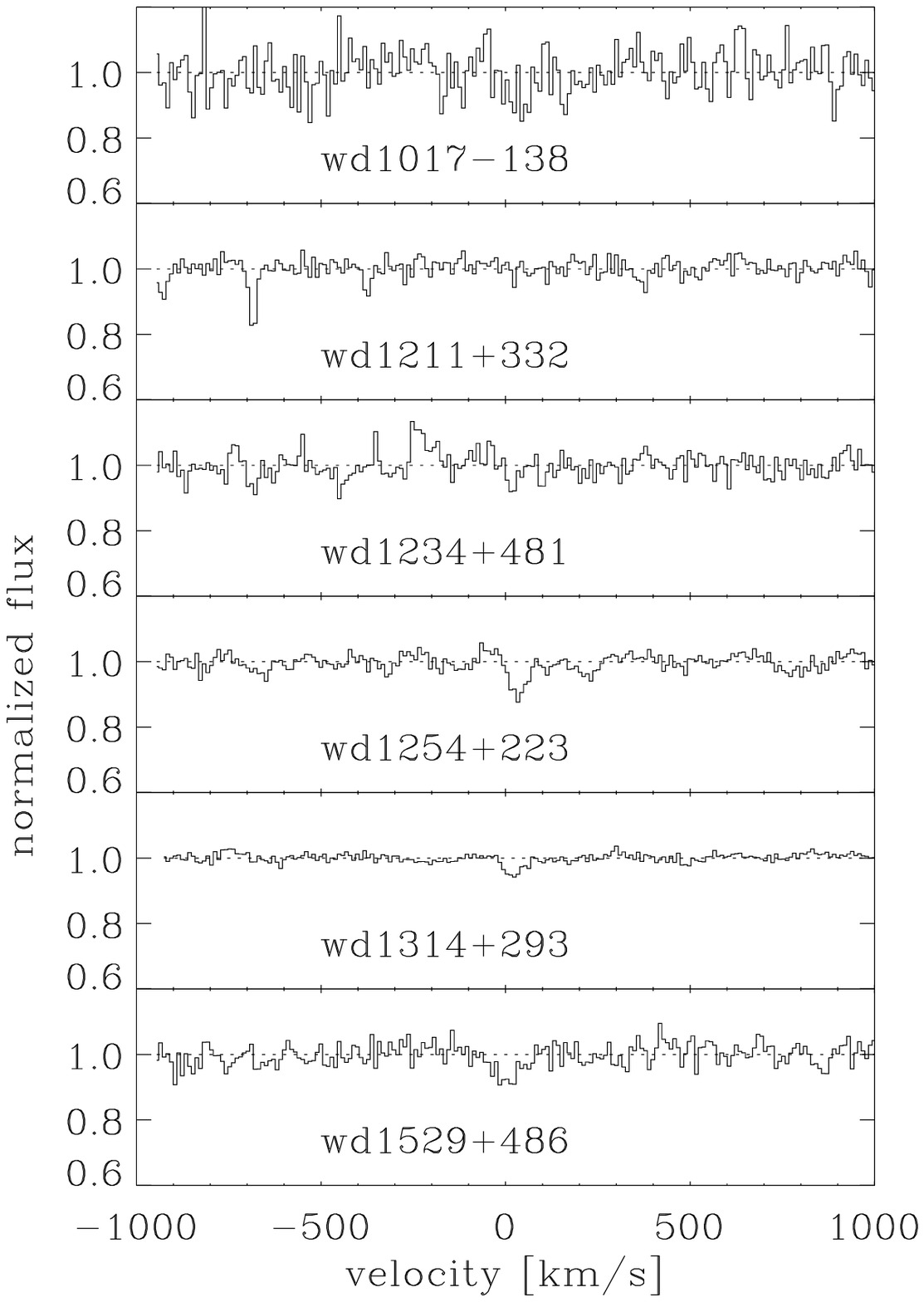}
\end{figure*}

\begin{figure*} 
\plotone{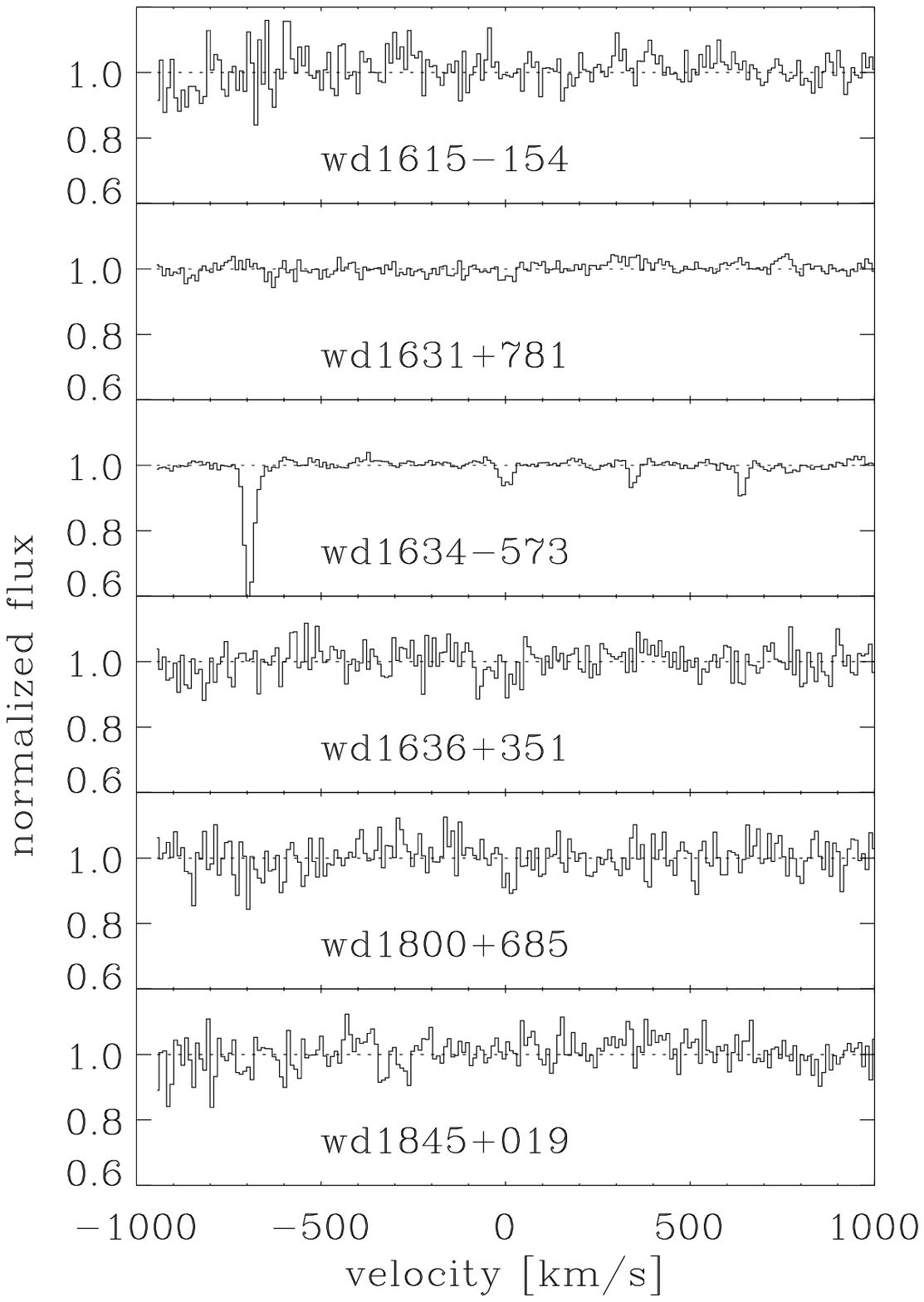}
\end{figure*}

\begin{figure*} 
\plotone{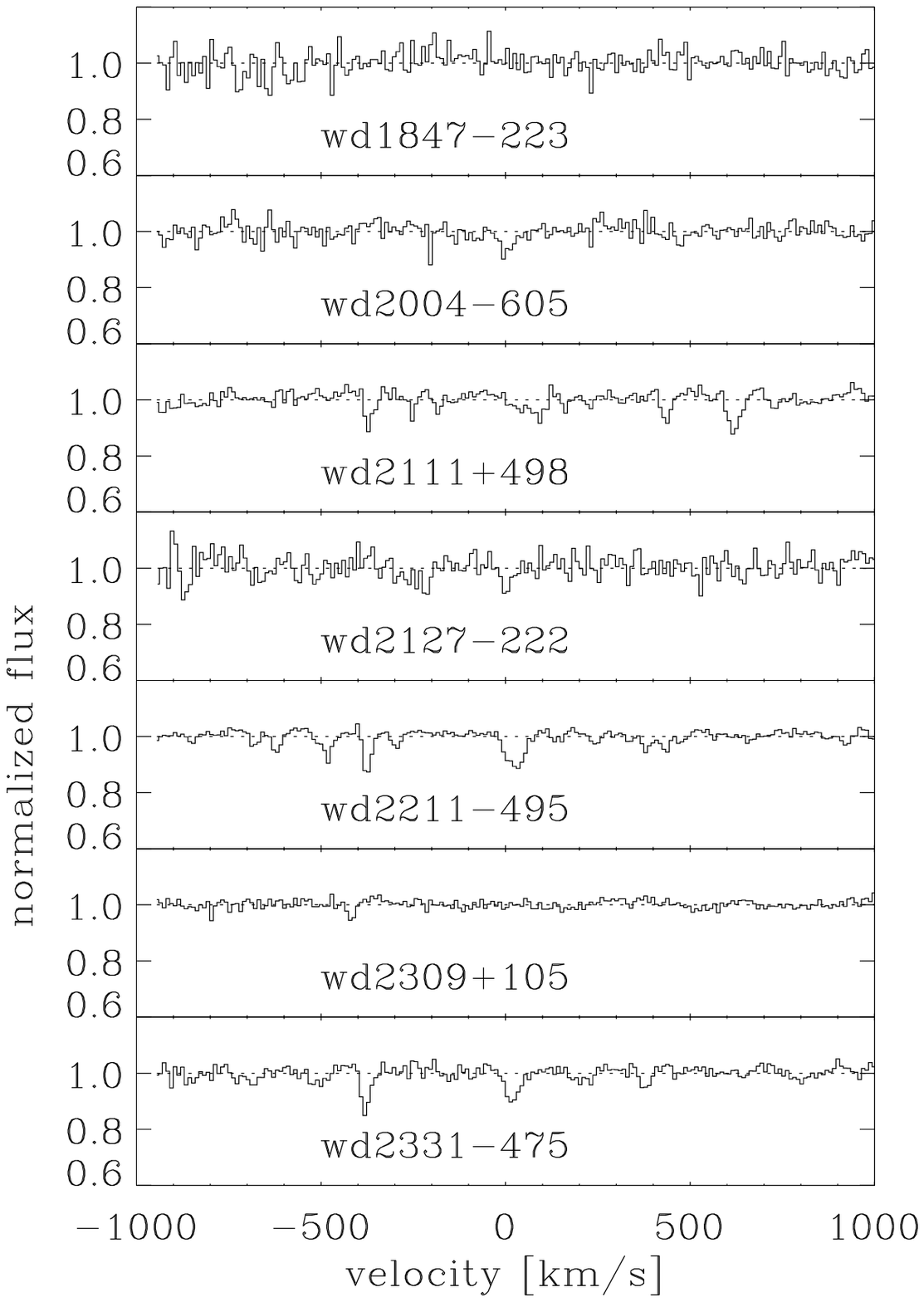}
\end{figure*}

\begin{figure*}  
\epsscale{1.0}   
\plotone{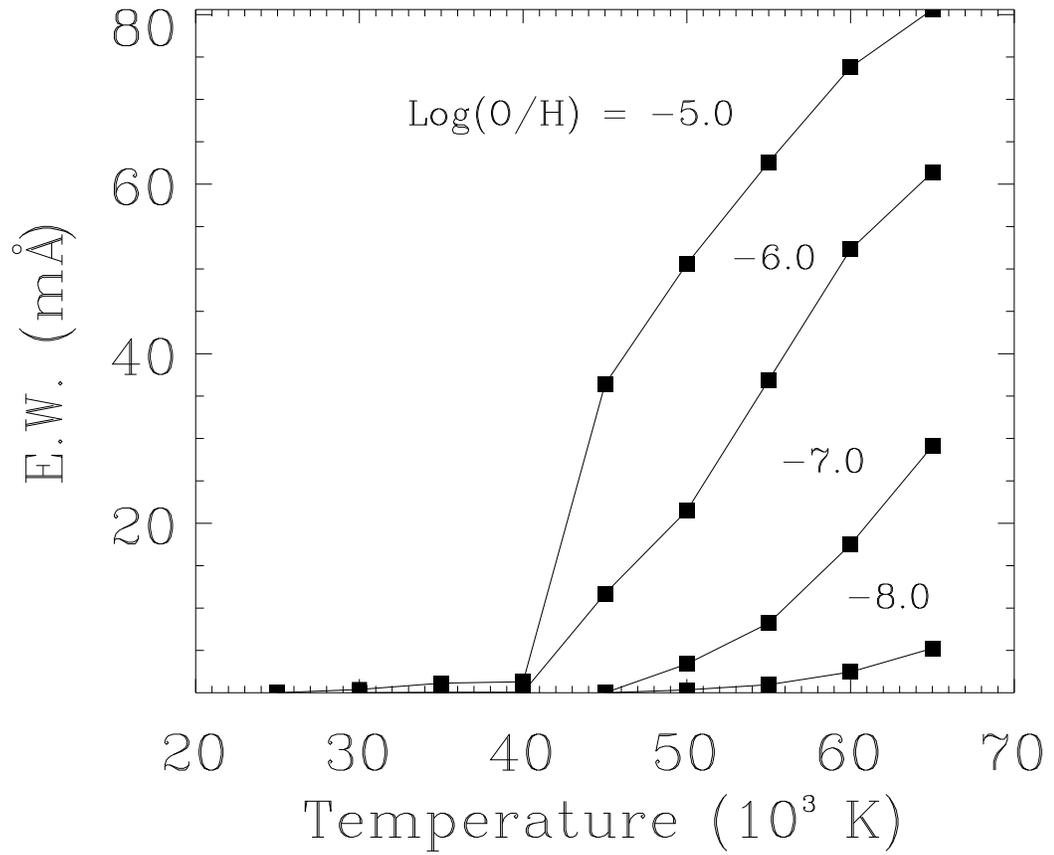}
\caption{Equivalent width of the \OVI\ \lam1031.9 photospheric 
line vs. T$_{\rm{eff}}$ for a grid of oxygen abundances, log(O/H), based
on NLTE white dwarf model atmospheres with $\log g = 8$.}
\end{figure*}

\begin{figure*}  
\epsscale{0.7}   
\plotone{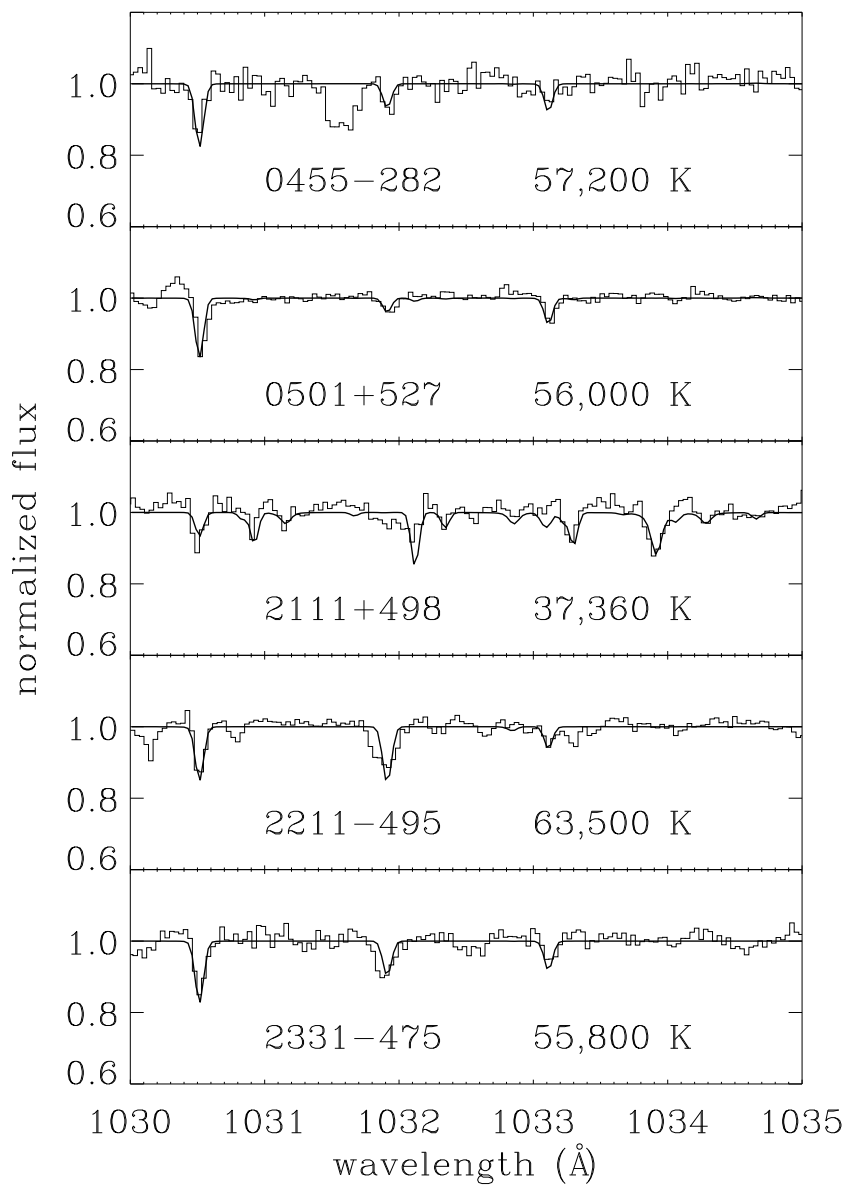}
\caption{Plots of 5 white dwarf spectra containing metals in their atmospheres.
The data, shown as a histogram plot, is the same as that in Figure 4, but
the wavelength zero-point has been shifted to put the 
stellar lines at their laboratory wavelengths. The solid dark curve overlayed
is a photospheric model.  The strongest
stellar lines visible here are \PIV\ \lam1030.51, \OVI\ \lam1031.93, 
and \PIV\ \lam1033.11.  A host of \FeIII\ photospheric lines are present 
in the spectrum of WD~2111$+$498, most notably the lines at 1030.92, 1032.12,
and 1033.30 \AA\/.}
\end{figure*}

\begin{figure*}
\epsscale{1.0}
\plotone{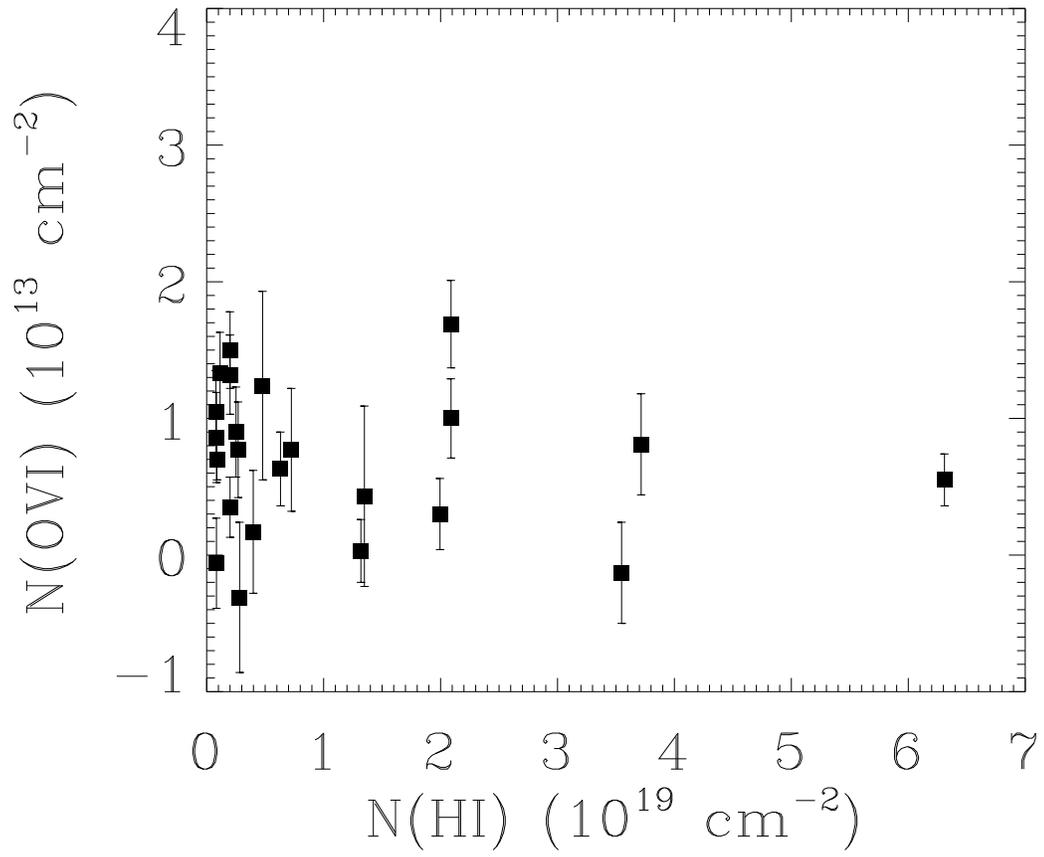}
\caption{The column density of \OVI\ versus \HI\/.  }
\end{figure*}

\begin{figure*}
\plotone{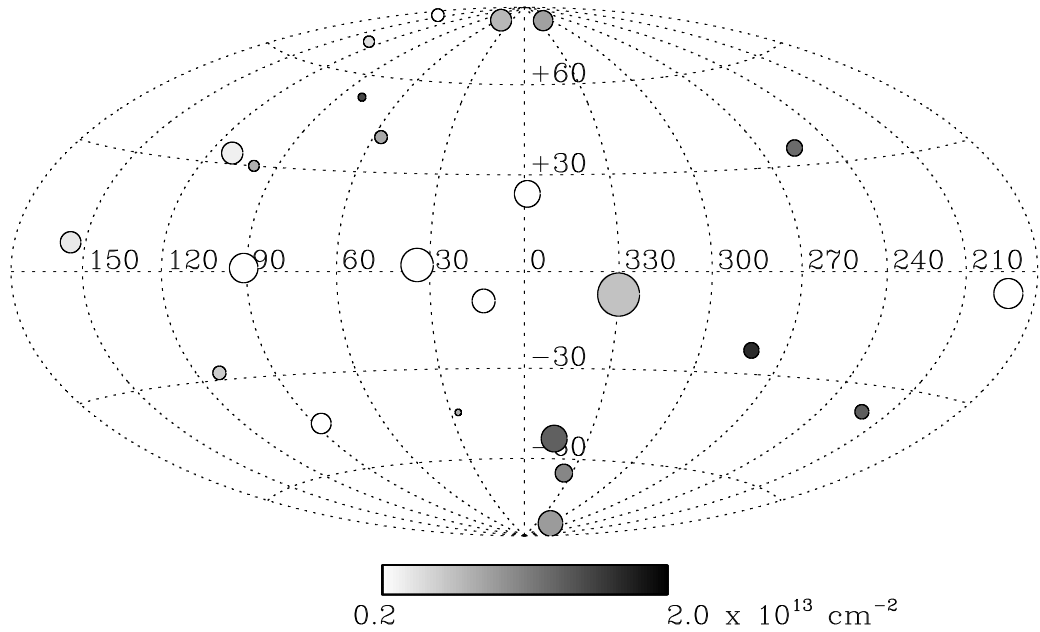}
\caption{The distribution of \OVI\ column density is plotted on an Aitoff
projection of the sky. Each of the 25 stars in the survey are plotted as
a shaded circle.  The diameter of the circle is inversely proportional to
the star's distance, with a diameter of 5 degrees corresponding to a 
distance of 50 pc.  The gray-scale shading is proportional to the \OVI\
column density along that line of sight, with a scaling as shown in
the color bar.  
}
\end{figure*}

\begin{figure*}
\plotone{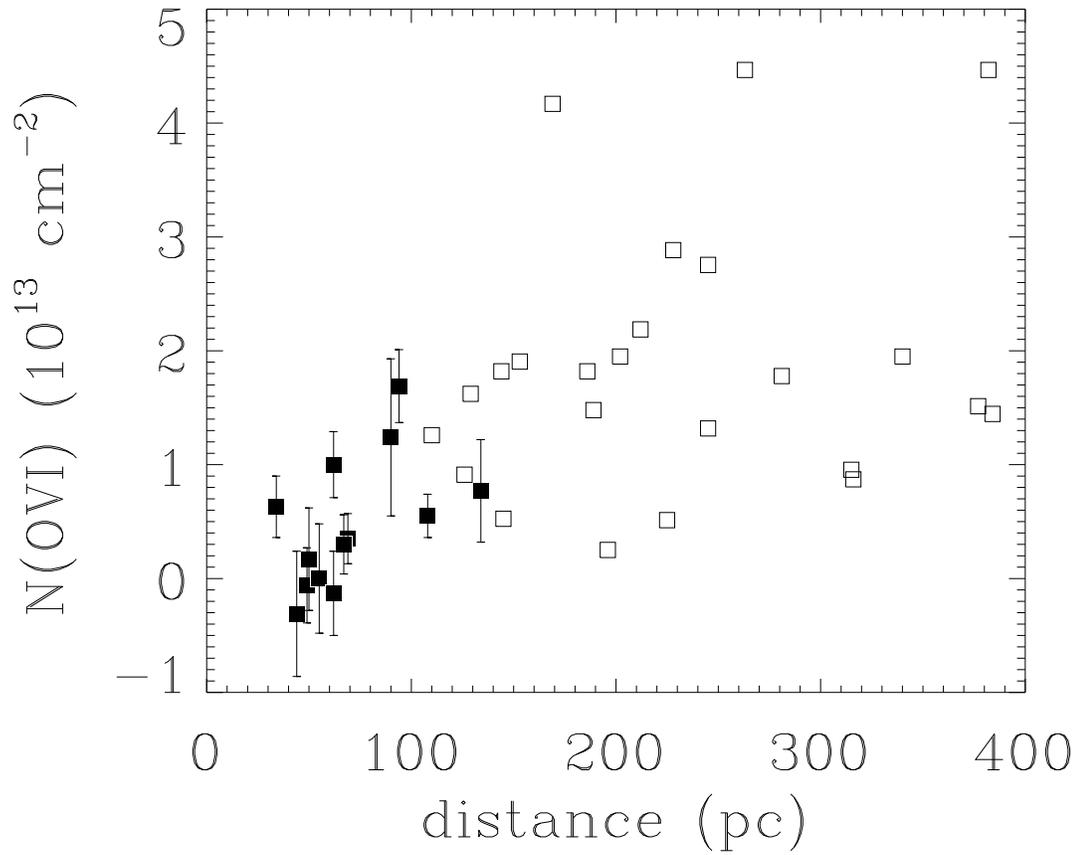}
\caption{The column density of \OVI\ versus stellar distance for stars
at Galactic latitudes $|b|<35^\circ$.  The solid squares
are the data for the {\it FUSE} LISM targets reported in this paper. The open
squares are column densities from {\it Copernicus} data, for stars at 
distances of $100-400$ pc \citep{jenkins78a}.}
\end{figure*}

\end{document}